\documentclass[aps,prd,reprint,twocolumn,preprintnumbers,floatfix,nofootinbib]{revtex4-1}
\usepackage{graphicx}
\usepackage{amsmath}
\usepackage{amsfonts}
\usepackage{ulem}
\usepackage{multirow}
\usepackage{placeins}
\usepackage{mathtools}
\usepackage{todonotes}
\usepackage{physics}
\usepackage{amssymb}
\usepackage{bm}
\usepackage{csquotes}
\usepackage{epstopdf}
\usepackage{float}
\usepackage{xcolor}

\usepackage{capt-of}
\allowdisplaybreaks
\usepackage{natbib}
\usepackage{hyperref}

\hypersetup{
     colorlinks   = true,
     citecolor    = red,
     linkcolor    = blue,
     urlcolor     = blue,
}
\definecolor{deeppink}{rgb}{0.9, 0.17, 0.31}

\def\ae{a_{\rm e}}
\def\He{H_{\rm e}}
\def\ke{k_{\rm e}}
\def\kre{k_{\rm RH}}
\def\te{t_{\rm e}}
\def\etae{\eta_{\rm e}}

\def\wre{w_{\phi}}

\graphicspath{{./figs/}}
\newcommand{\beq}{\begin{equation}}
\newcommand{\eeq}{\end{equation}}
\newcommand{\bea}{\begin{eqnarray}}
\newcommand{\eea}{\end{eqnarray}}
\newcommand{\Trh}{T_{\rm RH}}

\begin{document}

\title{
Generalizing the Bogoliubov vs Boltzmann approaches in gravitational production}
\author{Ayan Chakraborty$^{a}$}%
\email{chakrabo@iitg.ac.in}
\author{Simon Cléry$^{b, c}$}
\email{simon.clery@tum.de}
\author{Md Riajul Haque$^{d}$}
\email{riaj.0009@gmail.com}
\author{Debaprasad Maity$^{a}$}
\email{debu@iitg.ac.in}
\author{Yann Mambrini$^{b}$}
\email{yann.mambrini@ijclab.in2p3.fr}
\affiliation{%
${}^a$
	Department of Physics, Indian Institute of Technology Guwahati, Guwahati, Assam, India 
}%
\affiliation{
${}^b$ Universit\'e Paris-Saclay, CNRS/IN2P3, IJCLab, 91405 Orsay, France
 }
 \affiliation{${}^c$ Technical University of Munich (TUM), School of Natural Sciences, Physics Department, James-Franck-Str. 1, 85748 Garching, Germany,
 }
 \affiliation{
${}^d$
\,Physics and Applied Mathematics Unit, Indian Statistical Institute, 203 B.T. Road, Kolkata 700108, India
}

\newpage
\thispagestyle{empty}
\newpage
  
\begin{abstract}
We investigate the spectral behavior of scalar fluctuations generated by gravity during inflation and the subsequent reheating phase. We consider a non-perturbative Bogoliubov treatment within the context of pure gravitational reheating. We compute both long and short-wavelength spectra, first for a massless scalar field, revealing that the spectral index in part of the infrared (IR) regime varies between $-6$ and $-3$, depending on the post-inflationary equation of state (EoS), $0\leq w_\phi\leq1$. Furthermore, we study the mass-breaking effect of the IR spectrum by including the finite mass, $m_{\chi}$, of the daughter scalar field. We show that for $m_{\chi}/H_{\rm e} \gtrsim 3/2$, where $H_{\rm e}$ is the Hubble parameter during inflation, the IR spectrum of scalar fluctuations experiences exponential mass suppression, while for smaller masses, $m_{\chi}/H_{\rm e}<3/2$, the spectrum remains flat in the IR regime regardless of the post-inflationary EoS. For any general EoS, we also compute a specific IR scale, $k_m$, of fluctuations below which the IR spectrum will suffer from this finite mass effect. In the UV regime, oscillations of the inflaton background lead to interference terms that explain the high-frequency oscillations in the spectrum. Interestingly, we find that for any EoS, $1/9 \lesssim w_\phi \lesssim 1$, the spectral behavior turns out to be independent of the EoS, with a spectral index $ -6$. We have compared this Bogoliubov treatment for the UV regime to perturbative computations with solutions to the Boltzmann equation and found an agreement between the two approaches for any EoS, $0 \lesssim w_\phi \lesssim 1$. We also explore the relationship between the gravitational reheating temperature and the reheating EoS employing the non-perturbative analytic approach, finding that 
reheating can occur for $w_\phi \gtrsim 0.6$.
\end{abstract}

\maketitle

\section{\textbf{ Introduction }}\label{sec1}
Inflation has been proven to be a unique paradigm which has 
undergone a large number of cosmological tests \cite{Olive:1989nu,Linde:1990flp,Lyth:1998xn,Martin:2013tda,Martin:2013nzq,Martin:2015dha}. Starting 
from inflation, the Universe has gone through different phases 
which are yet to be understood and are subject to intensive 
investigations. During those 
phases, one of the important 
physical phenomena that garnered special interest 
recently is the quantum mechanical production of 
elementary particles, their impacts, and observability in 
the early Universe.

In the meantime, inflation should come to an end, and
it is well known that after the end of 
inflation \cite{Guth:1980zm,Linde:1981mu,Albrecht:1982wi,Starobinsky:1980te, Lemoine:2008zz, martin2003inflation, Martin, sriramkumar2009introduction, linde2014inflationarycosmologyplanck2013, Bassett_2006,  Piattella:2018hvi, Baumann_2009, Baumann:2018muz}, 
quantum fluctuations over the coherently 
oscillating inflaton background starts getting amplified because of their non-trivial coupling, and that can finally populate the universe 
which we observe today. 
This phenomenon can be interpreted as an energy transfer from the oscillation to particles.
This post-inflationary 
era is known as \textit{reheating}, and the reader can refer to
\cite{Kolb:1990vq,Shtanov:1994ce,Kofman:1997yn} among the earliest studies on the subject as well as to \cite{Allahverdi:2010xz,Sato:2015dga,Lozanov:2019jxc} for more recent reviews. It 
is a phase during which such particle production may 
eventually lead to a universe dominated by relativistic 
degrees of freedom, and traces the beginning of the radiation-
dominated phase\footnote{We will restrict ourselves in gravitational-type couplings, and we will thus neglect the effects due to the 
preheating phase \cite{Garcia:2021iag}, or fragmentation \cite{Garcia:2023dyf,Garcia:2024zir,Garcia:2023eol} } \cite{Garcia:2020wiy,Garcia:2020eof}. 

Several different production mechanisms 
have been proposed and studied quite extensively over the 
years. Gravitational production is one of them,
and it has recently re-attracted attention with the development of a novel framework of perturbative gravitational portals during reheating \cite{Bernal:2018qlk,Mambrini:2021zpp,Clery:2021bwz} and through several studies on the gravitational production of perturbations of different spins \cite{Ahmed:2020fhc,Kaneta:2023uwi,Choi:2024ilx}. Indeed, the main
interest of such a mechanism is that it does not require any 
new coupling parameter except gravity. It can then be considered as a minimal production, and {\it should} be added to any extension of the Standard Model. 
Particularly, the 
gravitational background dynamics can
be sufficient to reheat the universe \cite{Barman:2022qgt, Haque:2022kez, Haque:2023yra, Barman:2023opy}. The intriguing 
possibility of quantum mechanical particle production 
by the dynamical space-time was 
first put forward by Parker (1969) \cite{PhysRev.183.1057,PhysRevD.3.346}, and then thoroughly studied by many authors,
among them is L.H. Ford
who applied it in the framework of inflation.
In \cite{PhysRevD.35.2955,Ford:2021syk} he studied the particle creation 
due to the abrupt change of space-time metric during its 
transition from an inflationary era to a subsequent matter or 
radiation-dominated phase. Subsequently, extensive studies 
on this gravitational particle production has been performed 
and possibilities of reheating have also been analyzed\footnote{Note that already in 1939 \cite{SCHRODINGER1939899}, Schrodinger proposed to look at the particle-antiparticle transition in the Universe by solving his equation in a de Sitter space-time.} \cite{Dolgov:1989us,Ema:2016hlw,Ema:2018ucl,Chung:2018ayg,Basso:2021whd,Hashiba:2018iff,Hashiba:2018tbu,Herring:2019hbe,Lankinen:2019ifa,Pallis:2005bb}. 

While analyzing gravitational production, two commonly used approaches, namely \textit{Bogoliubov} and \textit{Boltzmann}, are employed to do a quantitative study of 
particle production, especially in the 
process of reheating\footnote{While preparing the submission, a paper \cite{feiteira2025cosmologicalgravitationalparticleproduction} appeared comparing two approaches for the computation of gravitational particle production from inflation. However, their paper considers only an inflaton with an EoS $w_\phi = 0$ and addresses solely the computation of the long-wavelength part of the spectrum (IR). The authors consider the Bogoliubov as well as the Starobinsky stochastic approaches, including a possible non-minimal coupling to gravity of the produced scalar field.}. \footnote{In \cite{Alexander:2024klf}, authors have studied the Bogoliubov transformation of parity-violating gravitational waves and spectral behavior across different equations of states. In their work, they have considered the transition from de Sitter inflation to the matter and radiation-dominated phase.} However, the Boltzmann approach is widely used, particularly concerning the production of dark matter {\it after} the period of inflation, 
and is based on solving the system of dynamical equations for different energy density components of the system under consideration. The crucial point is that the Boltzmann collision terms are calculated employing the {\it perturbative} QFT technique considering specific inflaton decay/annihilation channel through the exchange of a graviton
\cite{Garcia:2020eof,Moroi:2020bkq,Haque:2022kez,Haque:2021mab,Clery:2021bwz}. Therefore, such an approach typically deals with sub-horizon modes,
or {\it short-wavelength} modes, which means modes whose lower frequency corresponds (roughly) to the fundamental frequency of the inflaton at the end 
of inflation. By simple energy conservation principle, 
the 
Boltzmann technique does not include
the production of super-horizon modes also called 
{\it long-wavelength} modes,
like, for 
instance, quantum fluctuations of stable 
particles produced during inflation which 
re-enter the horizon during reheating, like spectator dark matter
\cite{Markkanen:2018gcw,Choi:2024bdn, Garcia:2025rut, Enqvist:2014zqa, Tenkanen:2019aij,Choi:2024bdn}. 
Further, it does not 
take into account the exact dynamics
of the background inflaton and treat 
interaction perturbatively. This approach can be considered as the one giving the minimal amount of matter (or radiation) produced, being ignorant of the inflationary process, even of its existence.~ 

On the other hand, the Bogoliubov approach is 
non-perturbative. The same system is solved explicitly in a given dynamical background 
keeping all the relevant interactions. Quantum 
particle production is studied through the 
evolution of quantum fields in classical background. The capability of capturing the minute details of background 
dynamics makes this non-perturbative treatment more complete compared to the perturbative one, {\it short-wavelength} components as well as {\it long-wavelength} ones being naturally 
included in the calculation.
Indeed, this approach naturally captures both super and sub-horizon modes, and that is found to be important, as we will discuss in detail
in this work.

We are not the first to deal with the comparison between the two 
techniques of gravitational production. Indeed, in a 
recent very interesting work \cite{Kaneta:2022gug}, the authors have shown an equivalence between the \textit{Bogoliubov} and \textit{Boltzmann} approaches.
However, they limited themselves 
to the case when the background dynamics is governed by the matter-like equation of state (EoS) $w_{\phi} \simeq 0$. 
In the present work, we extend the analysis further, and show that, in general, the two approaches yield quantitatively very 
different results for the spectrum
particularly for higher values of $w_\phi$.
We then explore 
the dynamics of reheating 
through this non-perturbative approach 
which has not been 
discussed in the 
literature in this general setup. In this 
paper we focus 
exclusively on the pure 
gravitational 
interaction, 
generalization to non-minimal coupling will be considered in future work.\\  
Pure gravitational reheating, on the other hand, has already been studied in the Boltzmann framework with significant details
\cite{Barman:2024slw,Garcia:2023obw,Kaneta:2023uwi,Barman:2022qgt,Clery:2022wib,Mambrini:2021zpp,Clery:2021bwz, Haque:2023zhb, Haque:2024zdq}. The unique feature of the mechanism lies in the fact that once we 
fixed a particular model of inflation, the only 
parameter that regulates the reheating dynamics is 
the inflaton EoS, $w_{\phi}$. The constraints are then very strong.
With the variation of $w_\phi$ from $0.65\to 0.99$ for $\alpha-$attractor model of inflation, the reheating temperature ($\Trh$) lies within BBN energy scale (i.e., $4$ MeV) to $10^{6}$ GeV \cite{Haque:2022kez,Clery:2021bwz}.
Within 
the same framework, the gravitationally produced scalar DM  was further restricted to be very light (between 0.1 eV and $\sim 1$ keV \cite {Clery:2021bwz,Haque:2021mab,Haque:2022kez}), and thus excluded by Lyman-$\alpha$ constraints.\\

We propose in this work to reanalyze the gravitational reheating scenario considering
the perturbative (Boltzmann)
and
non-perturbative (Bogoliubov) approach, 
taking into account
for the first time, the precise background dynamics. 
In particular, we investigate the qualitative and 
a quantitative difference arising between the two approaches in terms of distribution function and its predictions. 
We show that for super-horizon modes, the non-perturbative Bogoliubov approach yields distinct 
spectral behavior of the daughter field for which 
the Boltzmann analysis is not applicable. 
Moreover, 
our results indicate that the Bogoliubov
method predicts slightly higher 
reheating temperature 
for a given equation of state than the Boltzmann one. 

Nonetheless, through this submission, we want to advocate that to have a correct prediction, if the inflaton coupling parameters are within the perturbative regime, the Bogoliubov approach can capture the finer aspects of the physics of reheating, which has not been much explored in the literature.\\

The paper is organized as follows. In Section \ref{sec2}, we remind 
the general formalism of non-perturbative 
production in the more general case,  considering a scalar field non-minimally coupled to
gravity and also coupled to the inflaton with quartic interaction. 
We then show what the mode equations are in the special case of pure gravitational production. 
In Section \ref{sec3}, we study the post-inflationary dynamics of the oscillatory inflaton background introducing the post-inflationary oscillatory Hubble scale for the general inflaton equation of state. In Section \ref{sec4}, we present an extensive non-perturbative analytic approach for the sub and super-horizon spectrum of the scalar fluctuation for general reheating EoS. In Section \ref{sec5}, we first compute the Boltzmann distribution function of the fluctuation and then compare it qualitatively and quantitatively with the sub-horizon spectrum of the Bogoliubov approach. In section \ref{sec6}, using the sub and super-horizon spectra of the produced fluctuation, we study the pure gravitational reheating dynamics in a non-perturbative approach. Finally, Section \ref{sec7} concludes the paper by briefly stating the important findings of this work.   

\section{Non-perturbative production of scalar fluctuations: General Formalism}\label{sec2}
We first discuss the general non-perturbative formalism of particle production. We consider the following Lagrangian for the inflaton ($\phi$) and a massive scalar daughter field $(\chi)$ non-minimally coupled to gravity, in the Jordan frame\footnote{The metric signature is chosen to be (+, -, -, -).},  
\bea
&& \mathcal{L}_{[\phi,\chi]}=\sqrt{-g}
    \left[-\frac{M_P^2}{2}R+\frac{1}{2} \partial_{\mu}\phi \partial^{\mu}\phi-V(\phi)\right.
    \nonumber
    \\
    &&
    \left.+\frac{1}{2}\partial_\mu \chi \partial^{\mu}\chi-\frac{1}{2}(m_{\chi}^2+g^2\phi^2+\xi R)\chi^2\right]\,,
    \label{faction}
\eea
where the background FLRW metric is expressed as $ds^2=dt^2-a^2(t)d\vec{x}^2=a^2(\eta)\big(d\eta^2-d\vec{x}^2\big)$ 
with $a$ being the scale factor and $\sqrt{-g}=a^4(\eta)$. $V(\phi)$ is the inflaton 
potential,  $m_{\chi}^2$ is the bare mass of the produced inflaton
quanta, and $M_{P}={1}/{\sqrt{8\pi G}}\approx 2.435\times 10^{18}$ GeV is the reduced Planck mass. $\xi$ is 
the dimensionless non-minimal coupling of $\chi$ with gravity, and $g$ is 
a dimensionless coupling strength with the inflaton. 
The Ricci scalar $R$ generates a (time-dependent) effective mass for the $\chi$ field 
as, 
\beq
m_{\text{eff}}^2(\eta)=m_{\chi}^2+g^2\phi^2(\eta)+\xi R(\eta)\,.
\eeq
From the background inflaton part of the Lagrangian (\ref{faction}), 
the inflaton dynamical equation is given by
\begin{equation}\label{finflaton}
  \ddot \phi +3H\dot \phi+\pdv{ V(\phi)}{\phi}=0    \,,
\end{equation}
with the Hubble scale
\begin{equation}\label{Hubble}
 H=\sqrt{\frac{\frac{1}{2}\dot \phi^2+V(\phi)}{3M_{P}^2}}   \,.
\end{equation}
Expressing the scalar field $\chi$ in terms of Fourier modes,
\begin{equation}\label{ffouri}
     \chi(\eta,\vec{x})= \int\frac{d^3\vec{k}}{(2\pi)^{3/2}}  ~e^{i\vec{k}.\vec{x}}
     \left[\chi_{\vec{k}}(\eta)~a_k+\chi_{\vec k}^*(\eta) a^{\dagger}_{-k})\right]
 \end{equation}
 and using the Lagrangian  (\ref{faction}), we obtain the following equation for the mode function $\chi_{\vec{k}}$, 
   \begin{equation}\label{fdynamical1}       \chi_{\vec{k}}^{\prime\prime}+2\mathcal{H}\chi_{\vec{k}}^{\prime}+\left[k^2+a^2(\eta)\left(m_{\chi}^2+g^2\phi^2+\xi R\right)\right]\chi_{\vec{k}}=0\,,
  \end{equation}
where $\mathcal{H}={a^{\prime}}/{a}$, while the Ricci scalar is given by $R=-6 {a^{\prime\prime}}/{a^3}$, and all the derivatives are defined with respect to conformal time $\eta$. In Eq.~(\ref{fdynamical1}), we notice the presence of the damping term $2\mathcal{H} \chi_{\vec{k}}^{\prime}$ which is non-zero for an expanding cosmological background.
For convenience, it is conventional to rescale the field properly to absorb the damping term $2\mathcal{H} \chi_{\vec{k}}^{\prime}$, by defining the rescaled field 
\beq
X_{\vec{k}} = a\chi_{\vec{k}}\,,
\eeq 
and obtain the following mode equation,
   \begin{equation}\label{fdynamical2}       X_{\vec{k}}^{\prime\prime}+\bigg[k^2+a^2\big(m_{\chi}^2+g^2\phi^2\big)+\frac{a^2 R}{6}(1+6\xi)\bigg]X_{\vec{k}}=0\,.
  \end{equation}
  The bracketed term in the equation above    can be identified as a time-dependent frequency $\omega_k$ where
  \begin{equation}\label{ffrequency}      \omega_k^2(\eta)=k^2+a^2\big(m_{\chi}^2+g^2\phi^2\big)+\frac{a^2 R}{6}(1+6\xi)\,.
  \end{equation}

The time dependence in $\omega_k(\eta)$ renders impossible to decompose the scalar field $\chi(\eta,x)$ in
Eq.~(\ref{ffouri}) into a classical positive/negative frequency basis, with 
\beq
\chi(\eta)=\frac{e^{-i \omega_k \eta}}{\sqrt{2 \omega_k}}\,.
\eeq
In other words, the presence of a time-dependent background breaks the time translation symmetry, and no Killing vector field is
anymore associated to a positive
frequency eigenfunction. 
However, taking into account the evolution of the frequencies with time, one can still decompose $X(\eta,x)$ as
\bea
X(\eta,x)=
&&
\int\frac{d^3k}{(2 \pi)^\frac32}\frac{e^{i\vec k. \vec x}}{\sqrt{2 \omega_k(\eta)}}\left[
\left(\alpha_{\vec k}
e^{-i\Omega_k(\eta)}
+\beta_{\vec k}
e^{i\Omega_k(\eta)}\right)a_{\vec k}
\right.
\nonumber
\\
&&
\left.
+(\alpha_{\vec k}^*
e^{i\Omega_k(\eta)}+\beta_{\vec k}^*
e^{-i\Omega_k(\eta)})a_{-\vec k}^\dagger
\right]\,,
\label{Eq:Xx}
\eea
with 
\beq
\Omega_k(\eta)=\int^\eta\omega_k(\eta')d\eta'\,,
\eeq
and where $\alpha_{\vec{k}},~ \beta_{\vec{k}}$ are the time-dependent
Bogoliubov coefficients satisfying the normalization 
condition\footnote{Extracted from the Wronskian condition on $X(\eta,x)$, 
$X_k\,X_k^{*'}-X_k^* \,X_k'=i$.} $|\alpha_{\vec{k}}|^2-|\beta_{\vec{k}}|^2=1$. 
We can interpret this evolution in the mode $X_{\vec k}(\eta)$ 
 \begin{equation}\label{WKBform}
X_{\vec{k}}(\eta)=\frac{1}{\sqrt{2\omega_k}}\big[\alpha_{\vec{k}}e^{-i\int^{\eta}\omega_k(\eta^{\prime})d\eta^{\prime}}+\beta_{\vec{k}}e^{i\int^{\eta}\omega_k(\eta^{\prime})d\eta^{\prime}}\big]\,,
\end{equation}
in terms of an evolution in the creation/annihilation operator space. Indeed, rewriting Eq.~(\ref{Eq:Xx}) as
\bea
&&
X(\eta,x)=\int\frac{d^3k}{(2 \pi)^\frac32}e^{i \vec k . \vec x} \times
\\
&&
\frac{e^{-i \Omega_k}}{\sqrt{2 \omega_k}}
\left[\alpha_{\vec k} a_{\vec k}+\beta^
*_{\vec k}a_{-\vec k}^\dagger\right]
+\frac{e^{i \Omega_k}}{\sqrt{2 \omega_k}}\left[
\beta_{\vec k}a_{\vec k}+\alpha_{\vec k}a^\dagger_{-\vec k}\right]\,,
\nonumber
\eea
we can define a new set of operators $\tilde a_{\vec k}$
and $\tilde a^\dagger_{-\vec k}$ by
\beq
\tilde a_{\vec k}=\alpha_{\vec k} a_{\vec k}+\beta^*_{\vec k} a^\dagger_{-\vec k}\,,
~~ 
\tilde a^\dagger_{-\vec k}=\alpha^*_{\vec k}a^\dagger_{-\vec k}+\beta_{\vec k}a_{\vec k}\,.
\eeq
Within this new set of operators, we deduce from the number operator defined by
\beq
N = \int \frac{d^3k}{(2 \pi)^3} \tilde a^\dagger_{-\vec k} \tilde a_{\vec k}\,,
\eeq
that
\beq
\langle 0 |N |0\rangle=a^3 n_\chi=
\int\frac{d^3k}{(2 \pi)^3}
\langle 0 | \tilde a_{-\vec k}^\dagger \tilde a_{\vec k}|0\rangle=
\int \frac{d^3k}{(2 \pi)^3}
|\beta_k|^2\,,
\eeq
where we defined the initial 
vacuum at time $\eta \rightarrow -\infty$ by $a_{\vec k}|0\rangle=0$, or
$\beta_{\vec k}|_{\eta \rightarrow -\infty} \rightarrow 0$.
Note also that the occupation number, $|\beta_k|^2$, is equivalent
to a distribution function $f_\chi(|k|,t)$ in the Boltzmann
approach. The number density 
of particles $n_k$ with momentum $k$ can also be obtained with the solution (\ref{WKBform}) noting that \cite{Kofman:1997yn, Garcia:2022vwm} 
 \begin{equation}\label{fnoden}
   n_k= |\beta_{\vec k}|^2= \frac{1}{2\omega_k}|\omega_kX_{\vec k}-iX_{\vec k}^{\prime}|^2\,.
 \end{equation}
The whole subject of gravitational particle production is then summarized into computing the value of $\beta_k$ in different regimes.

Integrating Eq.~(\ref{fnoden}) over all momentum modes we get the total number and UV convergent energy density as \cite{PhysRevD.9.341, PhysRevD.36.2963, Garcia:2022vwm,  Kolb:2023ydq}
\begin{align}\label{fnoEden}
 n_{\chi}&=\frac{1}{a^3} \int \frac{d^3k}{(2 \pi)^3} n_k, \nonumber\\
 \rho_{\chi}&=\frac{1}{(2\pi)^3a^4} \int d^3k\omega_k n_k .
\end{align}
The general formalism that we have constructed in this section is the main foundation of the non-perturbative study of particle production. In order to get particle number density spectrum, first we need to solve Eq. (\ref{fdynamical2}) and then implementing the solution in (\ref{fnoden}) we obtain the number density spectrum of produced particles.

In this work, our primary focus concerns pure gravitational production. Therefore, in the forthcoming discussion, we shall ignore all the terms involving external coupling parameters like $\xi$ or $g$, which have been introduced in the generic Lagrangian (\ref{faction}). The mode equation (\ref{fdynamical2}) becomes,
\begin{equation}\label{fdynamical3}  X_{\vec{k}}^{\prime\prime}+\bigg[k^2+a^2m_{\chi}^2+\frac{a^2R}{6}\bigg]X_{\vec{k}}=0\,,
\end{equation}
with 
\beq
R=-6\frac{a''}{a^3}\,.
\eeq
The time-dependent frequency in the case of a pure gravitational case can be obtained from Eq. (\ref{ffrequency}) setting $\xi=0$ and $g=0$, which gives \footnote{Note that the oscillation of $R$ causes $ \omega_k^2$ to cross zero, thereby inducing a tachyonic instability, particularly for low-momentum modes(IR modes) \cite{Wang:2024lva}, we ignore such effect. }
\begin{align}
\omega_k^2(\eta)&=\bigg(k^2+a^2m_{\chi}^2+\frac{a^2R}{6}\bigg)\nonumber\\
&=\bigg(k^2+a^2m_{\chi}^2-\frac{a^{\prime\prime}}{a}\bigg)\nonumber
\\
&=\bigg(k^2+a^2m_{\chi}^2+\left(\frac{3 w_\phi-1}{2}\right)a^2H^2\bigg)
\label{Eq:omegakH}
\,,
\end{align}
where the last term, $\frac{a^{\prime\prime}}{a}$, carries all the information concerning the background dynamics and acts as a source term in the production process \cite{Markkanen:2015xuw,Herranen:2015ima}. 
In Eq.~(\ref{Eq:omegakH}), $w_\phi$ represents the equation of state of the dominant field, the inflaton, or pressure $P_\phi$, with $w_\phi=P_\phi/\rho_\phi$.
Note that, in the case of a de Sitter background with a Hubble rate\footnote{In this case, one can show that $\eta\simeq \eta_{\rm e} -\frac{1}{a\He}$, where the subscript \enquote{e} means the end of inflation.} $\He$, $\frac{a''}{a}=2 a^2 \He^2$ ($w_\phi=-1$), and
\beq
\omega_k^2(\eta)=a^2\left(\frac{k^2}{a^2} +m_\chi^2-2 \He^2\right)\,.
\label{Eq:omegak}
\eeq
From Eq.~(\ref{Eq:omegak}), we see that, during inflation, modes exit the horizon, and are frozen when $\frac{k}{a}<\sqrt{2} \He$, and (or) if $m_\chi<\sqrt{2}\He$. Then, after the end of inflation, $aH(a)$ decreases with time, modes reenter in the horizon.

\section{Post-inflationary background dynamics}\label{sec3}

After inflation, the Universe enters in a phase 
dominated by the oscillations of the inflaton field, 
and this plays an important role in 
the gravitational production
in the post-inflationary era. However, before computing the spectra of the fluctuations, we need to study the background dynamics considering a specific inflationary 
setup. 
In the present study, we shall consider one of the observationally favored models \cite{Kallosh:2013yoa, Martin:2013tda, Kallosh:2019hzo, Mishra:2024axb}, the $\alpha$-attractor E-model, whose potential takes the form 
\begin{equation}\label{alphaattractor}
V(\phi)=\Lambda^4\left(1-e^{-\sqrt{\frac{2}{3\alpha}}.\frac{\phi}{M_{P}}}\right)^{2n} \,.
\end{equation}
By varying the exponent $n$, we can achieve different power law forms of the potential. 

On the other hand, the amplitude of the potential $\Lambda$, which measures the energy content in the inflaton condensate, is 
constrained by the CMB measurement, and is related to the scalar spectral 
index $n_s$, the amplitude of the inflaton fluctuations measured as CMB normalization $A_s = 2.1\times 10^{-9}$, and the tensor to scalar ratio $r$. The model is favored by the latest \textit{Planck}, ACT, DESI, and
 BICEP/Keck combined (P+ACT+LB+BK18) observational data sets (see the references \cite{ACT:2025tim, ACT:2025fju, haque2025actdr6insightsinflationary, mondal2025constrainingreheatingtemperatureinflatonsm, Chakraborty:2025oyj}),
 where $n_s=0.9743\pm 0.0034$ at $68\%$ C.L. and $95\%$ C.L. upper limit on tensor-to-scalar ratio $r_{0.05}$ is obtained as $r_{0.05}<0.038$.
The  
parameter $\alpha$ in $V(\phi)$ determines the shape of the potential. The energy scale 
of inflation related to the parameter $\Lambda$ can be analytically expressed in terms of CMB parameters as \cite{Drewes:2017fmn}
     \bea
     \label{Lambdavalue}
     &&
      \Lambda= M_{P}\left(\frac{3\pi^2rA_s}{2}\right)^{\frac{1}{4}}
      \\
      &&
      \times
\left[\frac{2n(1+2n)+\sqrt{4n^2+6\alpha(1+n)(1-n_s)}}{4n(1+n)}\right]^{\frac{n}{2}}\,,
\nonumber
   \eea
   where the tensor-to-scalar ratio is expressed in terms of CMB parameters as
   \[
   r= \frac{192\alpha n^2(1-n_s)^2}{\Big[4n+\sqrt{16n^2+24\alpha n(1-n_s)(1+n)}\Big]^2}\,.
   \]\\ 
   During inflation, the inflaton satisfies usual slow roll conditions
   set by one of the slow roll parameters 
   \beq
   \epsilon\propto \left(\frac{V'}{V}\right)^2\,,
   \eeq
  and at the end of inflation, it is set to be unity. 
  In the context of the present $\alpha$-attractor E-type potential model, we can derive an expression of the field value for general $n$ at the inflation end as
\begin{equation}\label{phiend}
      \phi_{\text{e}}=\sqrt{\frac{3\alpha}{2}}M_{P}\ln{\left(\frac{2n}{\sqrt{3\alpha}}+1\right)},
\end{equation}
 where \enquote{e} signifies the value at the end of inflation or at the beginning of reheating. Using this field amplitude in (\ref{alphaattractor}), we obtain the potential at the end of inflation 
 \begin{equation}\label{Vend}
    V_{\text{e}}=\Lambda^4\left(\frac{2n}{2n+\sqrt{3\alpha}}\right)^{2n} \,.
 \end{equation}
 \begin{figure}[t]
\begin{center}
\includegraphics[scale=0.45]{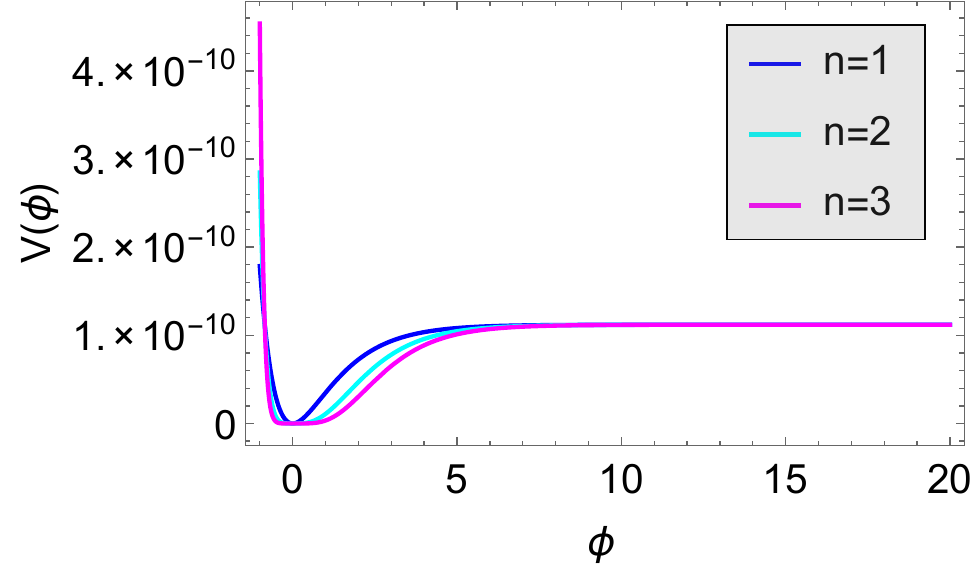}
\caption{\textit{Figure representing the variation of inflaton potential with the field, for $n=1,2,3$. Here $V(\phi)$ is in unit of $M_{P}^4$ and field $\phi$ is in $M_{P}$ unit. The flat portion of these potentials essentially indicates the inflationary era, and at the end of inflation, field $\phi$ starts oscillating around the potential minima near $\phi \sim 0$.}}
\label{threepot}
\end{center}
\end{figure}
 We illustrate in Fig.~(\ref{threepot}) the shape of the potential for different values of $n$. At the end of 
 the inflationary phase, inflaton field 
 starts oscillating around its minimum. 
 Expanding the potential (\ref{alphaattractor}) around this minimum,
 we obtain an approximated power-law potential
\begin{equation}\label{Vphin}
   V(\phi)\sim \Lambda^4\left(\frac{2}{3\alpha}\right)^{n}\left(\frac{\phi}{M_{P}}\right)^{2n}\,.
\end{equation}
In an expanding background, the amplitude of the inflaton field falls with time. We can 
then safely assume 
${\phi}/{M_{P}}\ll1$ to be true for a 
sufficient duration of the reheating phase. This justifies that we can simply neglect the higher-order terms 
in the expansion of the potential (\ref{alphaattractor}). In the post-inflationary era, inflaton field oscillates 
around the minima of this approximated power-law potential (\ref{Vphin}) with a decaying amplitude $\Phi(t)$. Oscillatory behavior is 
strongly influenced by the nature of the potential near its minimum. At this stage, the generic form of inflaton solution in the post-inflationary phase can be parameterized as
\beq
\phi(t)=\Phi(t) \mathcal{P}(t) = \Phi(t) \sum_{\nu\ne 0} \mathcal{P}_\nu e^{i\nu\omega t}\,,
\eeq
where $\Phi(t)$ is the inflaton 
amplitude and $\mathcal{P}(t)$ is quasi-periodic oscillatory function with the fundamental frequency $\omega$ \cite{Garcia:2020eof}.
The energy-density, $\rho_{\phi}$ and pressure, $P_{\phi}$, of homogeneous scalar field inflaton can be written as
\begin{equation}\label{avgpphirhophi}
   \rho_{\phi}=\frac{1}{2}\dot{\phi}^2+V(\phi), \quad \text{and} \quad P_{\phi}=\frac{1}{2}\dot{\phi}^2-V(\phi) \, .
\end{equation}
Having the approximated power-law potential (\ref{Vphin}), ignoring background expansion and taking oscillation average over one complete oscillation, we have average energy density $ \langle\rho_{\phi}\rangle=V(\Phi)$, and pressure as $\langle P_{\phi}\rangle=w_\phi \langle\rho_{\phi}\rangle$. The average post-inflationary EoS is then expressed as $w_{\phi}={(n-1)}/{(n+1)}$. In order to obtain the decaying amplitude $\Phi(t)$ for general $n$, we use (\ref{avgpphirhophi}) together with (\ref{Vphin}) and the form of average EoS $w_{\phi}$ in the continuity equation, $\langle\dot{\rho}_{\phi}\rangle +3H(1+w_{\phi})\langle\rho_{\phi} \rangle \simeq 0$, which leads to
\begin{align}\label{phiampli}
    \dot{\Phi}=-\frac{3H}{n+1}\Phi 
   ~~ \Rightarrow ~~\Phi(a)=\phi_{\text{e}}\left(\frac{a}{a_{\text{e}}}\right)^{-\frac{3}{n+1}}\,,
\end{align}
where $\phi_{\text{e}}$ and $a_{\text{e}}$ are inflaton amplitude and scale factor at the end of inflation, respectively.

In order to find the inflaton oscillation frequency $\omega$, using Eq.~(\ref{avgpphirhophi}) we write the dynamical equation for the quasi-periodic function,
\begin{align}\label{Pdynamical}
  &\dot{\mathcal{P}}\simeq -\frac{\sqrt{2(\rho_{\phi}-V(\phi))}}{\Phi}\simeq - \frac{m_{\phi}}{\sqrt{n(2n-1)}}\sqrt{1-\mathcal{P}^{2n}} \, ,
\end{align}
where we used $\rho_\phi=V(\Phi)$ and $\Phi(t)$ almost constant over one oscillation.
The effective inflaton mass $m_{\phi} =
  \sqrt{{\partial^2 V}/{\partial \phi^2}}$ is defined with respect to the envelope of the inflaton field  ${\Phi(t)}$
\begin{equation}\label{mphi}
    m_{\phi} =
  \sqrt{2n(2n-1)}\left(\frac{2}{3\alpha}\right)^{\frac{n}{2}}\left(\frac{\Lambda}{M_{P}}\right)^2\left(\frac{\Phi}{M_{P}}\right)^{(n-1)}M_{P} \, .
\end{equation}
From Eq.~(\ref{Pdynamical}), we deduce the fundamental background  oscillation frequency  \cite{Garcia:2020wiy}
\begin{equation}\label{fundafre}
    \omega=m_{\phi}\sqrt{\frac{\pi n}{(2n-1)}}\frac{\Gamma\left(\frac{1}{2}+\frac{1}{2n}\right)}{\Gamma\left(\frac{1}{2n}\right)} \,.
\end{equation}
Once the time-variation of inflaton amplitude is computed for general $n$, using (\ref{phiampli}) together with its oscillatory 
part $\mathcal{P}(t)$, we can express the leading order behavior of the 
Hubble scale in Eq.~(\ref{Hubble}) in terms of the decaying amplitude and quasi-periodic oscillatory function \cite{Chakraborty:2023lpr},
\begin{align}\label{Hubbleforn}
 H(a)\simeq \bar{H}\left(1+\frac{\mathcal{P}\sqrt{6(1-\mathcal{P}^{2n}})}{2(n+1)}\left(\frac{\phi_{\text{e}}}{M_{P}}\right) \left(\frac{a}{a_{\text{e}}}\right)^{-\frac{3}{n+1}}\right) \,,
\end{align}
where the first term is the slowly varying decaying term $\bar{H}= \He \left({a}/{a_{\text{e}}}\right)^{-\frac{3n}{n+1}}$ with  $H_{\text{e}}=\sqrt{{V_{\text{e}}}/{2M^2_{P}}}$ 
being the inflationary energy scale \footnote{The variation of the Hubble parameter is very small during any slow-roll model of inflation. For simplicity, we consider the de Sitter approximation; the Hubble parameter is constant throughout inflation.}
The second term inside the bracket can be identified with 
the fast varying oscillatory part in Hubble scale $H_{\text{fast}}$. We will see that this oscillation will play a crucial role in the process of particle production. 
Substituting the expression (\ref{phiampli}) in $H^2\simeq {V(\Phi)}/{3M_{P}^2}$, we obtain the leading order time variation of scale factor after inflation for general EoS as
\begin{equation}\label{scaleforn}
 a(t)=  a_{\text{e}} \left(\frac{t}{ t_{e}}\right)^{\frac{n+1}{3n}} \,,
\end{equation}
where 
\beq
t_{e}  = \frac{\sqrt{2n(2n-1)}(n+1)M_{P}}{m_{\phi}^{\text{e}}\sqrt{3}\,n\,\phi_{\text{e}}}\,,
\eeq
with
\beq
m_{\phi}^{\text{e}} 
  \simeq \sqrt{2n(2n-1)}\left(\frac{2}{3\alpha}\right)^{\frac{1}{2}}\frac{\Lambda^{\frac{2}{n}}}{M_{P}} \left(V_{\text{e}}\right)^{\frac{(n-1)}{2n}}\,.
  \label{mphiend}
\eeq

\section{Fluctuation spectrum for general reheating EoS ($w_{\phi}$), an analytical approach }\label{sec4}

The total energy density of produced scalar field fluctuations $\chi$ is 
computed by integrating the product of $|\beta_k|^2$ by the energy per momentum mode 
over all scales. Therefore, the nature of the $|\beta_k|^2$ spectrum 
should be thoroughly studied before proceeding toward the determination of the total energy of gravitationally 
produced fluctuations. This section will be 
devoted to the analytic determination of the spectrum in two different regimes. Indeed, the nature of the spectrum differs for sub-horizon modes ($k\gg a_{\text{e}}H_{\text{e}}\equiv \ke$) and super-horizon modes ($k\ll \ke$). For sub-horizon modes, the frequency term $\omega_k^2(\eta)$ in Eqs.~(\ref{Eq:omegakH}) and (\ref{Eq:omegak}) is always positive. Indeed, if at the end of inflation, 
$k > a H$, it will always be the case until present time if $w_\phi>-\frac13$, because $H$ decreases as $H\propto a^{-\frac{3(1+w_\phi)}{2}}$.
Therefore, 
those modes will not experience any tachyonic 
instability. However, as we will discuss later, naively the 
oscillatory behavior of the Hubble scale {\it during reheating} can still have a noticeable effect on those 
sub-horizon modes.
The density spectrum should depend on $w_\phi$ which sets the re-entry time of the mode $k$.
But, as we will see in our analysis, it will not be the case for $w_\phi>\frac19$.
On the other hand, modes satisfying $k^2<\left(\frac{a^{\prime\prime}}{a}-a^2m_{\chi}^2\right)|_{a_{\rm e}}=a_{\rm e}^2(2H_{\rm e}^2-m_\chi^2)$
causes frequency $\omega_k^2 (\eta)$ to become negative at a given time {\it during inflation } 
and this results  
tachyonic growth of those mode functions. 
Upon their horizon exit during inflation, the amplitude of those modes freezes until their re-entry during the post-inflationary evolution.
Thus, these super-horizon modes are not in the 
corresponding Bunch-Davies vacuum in post-inflationary evolution.
The dependence on $w_\phi$ should then be important to determine the time (scale factor) at re-entry.
To accurately track their evolution, one has to follow them from early times, when they were deep 
inside the horizon \cite{Garcia:2022vwm}, until their exit. Besides this, these super-horizon modes having 
very long-wavelength, do not get affected by the background oscillation of the inflaton during reheating.

In order to compute the full spectrum $|\beta_k|^2$,  we compute it separately, first for the super-horizon and then for the sub-horizon modes, as they 
experience different dynamics in two different regimes. We will then compare the total spectrum 
with the one obtained using a Boltzmann approach, where we compute the spectrum from 
Feynman amplitudes through the exchange of a graviton \cite{Mambrini:2021zpp,Clery:2021bwz}.

\subsection{Super-horizon modes spectrum $(k < \ae \He)$}

\subsubsection{Solution during the inflationary phase ($\eta < \eta_e$)}
The long-wavelength spectrum gets enhanced through tachyonic instability after horizon 
crossing during inflation. Depending upon different post-
inflationary EoS, this large-scale spectrum follows different power-law behavior during 
reheating. To take into account the spectrum's growth after the 
horizon crossing, we need to start studying its dynamics from the inflationary era itself.  
The evolution of scale factor for any $w_\phi$ can be represented as function of the conformal time as 
\beq
\label{scalefactor}
     a=
-\frac{1}{H_{\rm e}\eta_{\rm e}}\left(\frac{1+3w_{\phi}}{2|\eta_{\rm e}|}\right)^{\frac{2}{1+3w_{\phi}}}\left(\eta-\eta_{\text{e}}+\frac{2 |\eta_{\rm e}|}{1+3w_{\phi}}\right)^{\frac{2 }{1+3w_{\phi}}}  \,,
\eeq
with $\etae=-\frac{1}{H_{\rm e}a_{\rm e}}$, which gives 
during inflation ($w_\phi=-1$), 
\beq
a=-\frac{1}{\He \eta}\,,
\label{Eq:adesitter}
\eeq
valid in the corresponding range $\eta_i<\eta\leq \etae$.
Considering pure de Sitter inflation, for which $H=H_i=H_{\text{e}}$, 
it is trivial to check that during the transition from inflation to reheating, the scale 
factor and its first derivative change continuously at the junction point, that is at the end of inflation, $\eta=\etae$.

Violation of adiabaticity condition due to abrupt transition in the mode equation from the de Sitter phase to the 
oscillatory regime causes particle production observed upon modes re-entry associated with long-wavelength modes. Let's label $X_k^{(1)}(\eta)$ the adiabatic vacuum solution during the de Sitter phase in the time 
interval $\eta_i<\eta\leq\etae$, 
and $X_k^{(2)}(\eta)$ the associated vacuum solution during the reheating phase for 
$\eta\geq\etae$. Making these fields 
solutions and their first derivatives continuous at the junction $\eta=\etae$, the time-independent Bogoliubov coefficients $\beta_k$ associated with the number spectrum well into reheating era is obtained using \cite{deGarciaMaia:1993ck, Kaneta:2022gug, Kolb:2023ydq},
and following Eq.(\ref{WKBform})
\beq
X_k^{(1)}=\alpha_k X_k^{(2)}+\beta_k X_k^{(2)^*}\,,
\eeq
combined with the Wronskian condition
\beq
X_k^{(i)}X_k^{(i)^{*'}}-X_k^{(i)^*}X_k^{(i)'}=i\,,
\eeq
which gives the time-independent coefficient
\begin{equation}\label{bogo}
  \beta_k=i\left( {X_k^{(2)}}'(\etae) X_k^{(1)}(\etae) -{X_k^{(1)}}'(\etae) X_k^{(2)}(\etae) \right)\,.
  \end{equation}
To compute $\beta_k$ one then needs to determine $X_k^{(1)}(\eta)$ {\rm and} $X_k^{(2)}(\eta)$, 
solving the equation of motion before and after $\eta_e$.

During the de Sitter phase ($\eta\leq\eta_{\text{e}}$), 
the mode equation of a masslesss $\chi$-field for long-wavelength approximation, 
Eq.~(\ref{fdynamical3}) with $a(\eta)$ given by Eq.~(\ref{Eq:adesitter}), becomes
\begin{equation}\label{fregion2e}
    X_k^{\prime\prime}+\left[k^2-\frac{2}{\eta^2}\right] X_k=0 \,,
\end{equation}
which has for solution the Bunch-Davies mode function \cite{Baumann:2018muz,Riotto:2002yw}
\beq\label{fregion1}
  \boxed{ X_k^{(1)}(\eta)
   =\frac{e^{-ik \eta}}{\sqrt{2 k}}\left[1-\frac{i}{k \eta}\right]
   \simeq -\frac{i}{\sqrt{2}k^{\frac{3}{2}}\eta}e^{-ik\eta}} \,,
   \eeq
   and
   \beq
   {X_k^{(1)}}^{\prime}(\eta)=\frac{d X_k^{(1)}(\eta)}{d \eta}\simeq i\frac{e^{-ik\eta}}{\sqrt{2}k^\frac32\eta^2}\,,
\eeq
where we approximate $\frac{k}{a}\ll H_{\rm e}=-\frac{1}{a~\eta}$, or equivalently $|k\eta|\ll 1$.
The divergent term proportional to $\frac{1}{k \eta}$ is clearly
the signature of the dominant  (tachyonic) negative term $-\frac{a''}{a}$ in Eq.~(\ref{Eq:omegak}). In case of pure de Sitter inflation, one can also understand
that, at $\eta_{\rm e}$, the longer is the wavelength (the smaller is $k$), the sooner the mode enters the divergent regime. We expect then a power spectrum $\propto |X_k^{(1)}(\eta)|^2$ tilted toward the IR part of the spectrum. 

\subsubsection{Solution during the reheating phase ($\eta> \eta_e$)}

The same $k$-mode of massless scalar when passing through the reheating phase dominated by an equation of state $w_\phi$
(for $\eta\geq\eta_{\text{e}}$), after implementing  Eq.~(\ref{scalefactor}) in Eq.~(\ref{fdynamical3}), satisfies
\begin{equation}\label{fregion2e}
    X_k^{\prime\prime}+\Bigg[k^2
    -\frac{2(1-3w_\phi)}{(1+3w_\phi)^2}\frac{1}{\Big(\eta-3\bar \mu \eta_{\rm e}\Big)^2}\Bigg]X_k=0 \,,
\end{equation}
with
\beq
  \bar{\mu}=\frac{(1+w_{\phi})}{(1+3w_{\phi})}\,,
  \label{Eq:mubar}
\eeq
and whose solution is
\begin{equation}\label{fregion2s1}
   X_k(\eta)=C_1 2^{2\bar{\nu}}\Gamma(\bar{\nu}+1)\sqrt{2 i k\bar{\eta}} I_{\bar{\nu}}(i k \bar{\eta})+C_2\sqrt{\frac{2 i k\bar{\eta}}{\pi}}K_{\bar{\nu}}(i k\bar{\eta})\,,
\end{equation}
where $\bar{\eta} = \eta -{3\bar{\mu}}\eta_{\rm e}$, 
and $I_{\bar{\nu}}$, $K_{\bar{\nu}}$ are modified Bessel functions of order $\bar{\nu}$ with 
\begin{equation}\label{Eq:nubar}
    \bar{\nu} = \frac{3}{2}\frac{(1-w_\phi)}{(1+3w_\phi)}
\end{equation}
with $C_1$ and $C_2$ integration constants. We show in Table (\ref{tabspectralindex}) the values of $\bar \nu$ for a set of $\wre$.
\begin{table}[h!]
 \caption{\textit{Variation of the long-wavelength spectral indices \enquote{$\bar{\nu}$} with different EoS \enquote{$\wre$} }}
 \vspace{2ex}
 \centering
\begin{tabular}{||c|c| c||}
 \hline
 $n$ &$\wre$ & $ \bar{\nu}$ \\
 \hline\hline
 1 & 0 & 3/2\\ 
 \hline
 2 & 1/3& 1/2 \\
 \hline
 3 & 1/2 & 3/10 \\
 \hline
 4 & 3/5& 3/14 \\
 \hline
 5 & 2/3& 1/6 \\
 \hline
 6 & 5/7& 3/22 \\
 \hline
 7 & 3/4&  3/26 \\
 \hline
 9 & 4/5& 3/34\\
 \hline
 19 & 9/10& 3/74 \\ 
 \hline
 199 & 99/100& 3/794 \\
 \hline
\end{tabular}
\label{tabspectralindex}
\end{table}
Note that the generic solution obtained in Eq.~(\ref{fregion2s1}) could also have been applied 
to find the solution $X_k^{(1)}$ in the de Sitter phase.
Indeed, setting $w_\phi=-1$ in (\ref{fregion2s1}), 
$\bar \mu=0 \rightarrow \bar \eta = \eta$ and $\bar \nu=-3/2$,
limits of $\bar \nu$ for $w_\phi \rightarrow -1$.
We would have obtained
\beq
X_k^{(1)}(\eta)=\sqrt{\frac{\pi}{2k}}
\sqrt{2i k|\eta|}I_{-3/2}(i k |\eta|)\,,
\eeq
which gives $X_k^{(1)}(\eta)\sim -1/k^{3/2}\eta$ in the 
long-wavelength approximation, as obtained in Eq.~(\ref{fregion1}). 
We can also obtain the same expression (\ref{fregion1}) from the general solution in terms of the Hankel function, $X_k^{(1)}(\eta)=-i\sqrt{\frac{\pi\eta}{2}}H_{3/2}^{(1)}(k|\eta|)$. 
Therefore, these two forms of $X_k^{(1)}(\eta)$ in terms of $I_{-3/2}(i k |\eta|)$ and $H_{3/2}^{(1)}(k|\eta|)$ are 
equivalent to our solution (\ref{fregion1}). Note also that the equation (\ref{fregion2e}) is the same for
$w_\phi=0$ and $w_\phi=-1$, replacing $\eta$ by $\bar \eta$.
However, the solution for $w_\phi=0$ 
is {\it non-divergent} as $\frac{1}{k \bar \eta} \propto \frac{1}{\sqrt{a}}\rightarrow 0$ for increasing values of $a$.
Even if these remarks may seem too technical, we believe
they are important in the context, because several solutions,
functions and methods are given in the literature, without underlining their equivalence.

In order to compute the 
integration constants $C_1$ and $C_2$ for any value of $w_\phi$, we chose the \textit{adiabatic vacuum} for each 
mode. If spacetime changes very slowly, or equivalently, particle momentum is so 
large that it hardly feels the background dynamics, the mode function can be safely 
assumed to behave (and stay) as a positive frequency mode in Minkowski space in its asymptotic limit. 
This corresponds to the limit $k\eta\gg1$, and the mode solution (\ref{fregion2s1}) becomes
\begin{equation}\label{fregion2s2}
   X_k(\eta)\sim \Bigg[C_1\frac{2^{2\bar{\nu}}\Gamma(\bar{\nu}+1)}{\sqrt{\pi}}e^{ik \bar{\eta} } +C_2e^{-ik \bar{\eta}}\Bigg] .
\end{equation}
On the other hand, in the adiabatic vacuum limit, mode function becomes
\begin{equation}\label{adiabaticv}
    X_k(\eta)\xrightarrow{{\eta}\rightarrow \infty} \frac{e^{-ik\eta}}{\sqrt{2k}},
\end{equation}
where we used $\eta \rightarrow \infty \Rightarrow \bar \eta \rightarrow \infty$.
Comparing (\ref{fregion2s2}) with (\ref{adiabaticv}) we then deduce
\begin{equation}\label{c1c2}
    C_1=0, \quad
    C_2= 
    \frac{1}{\sqrt{2k}}e^{-3ik\bar \mu \eta_{\rm e}}=
     \frac{1}{\sqrt{2k}}e^{3i\bar \mu \frac{k}{\ke}}\,,
\end{equation}
where $\ke=-\frac{1}{\eta_{\rm e}}=\ae \He$ is the scale that left the horizon at the end of inflation.
Therefore, the adiabatic vacuum solution of massless particles for general EoS $w_{\phi}$ becomes
\begin{equation}\label{fregion2s3}
\boxed{
X_k^{(2)}(\eta)=\sqrt{\frac{\bar{\eta}}{\pi}}
e^{i\left(3\bar \mu \frac{k}{\ke}+\frac{\pi}{4}\right)}
\times
K_{\bar{\nu}}(i k \bar{\eta})}\,.
\end{equation}
Note that we recover the plane wave solution of the Minkowski space (Universe with an 
infinite pressure, or $w_\phi\rightarrow \infty$), corresponding to $\bar \nu=-\frac12$ and $\bar \mu=\frac13$, using $K_{-\frac12}(x)=\sqrt{\frac{\pi}{2x}}e^{-x}$,
as well as for $w_\phi=\frac13$ ($\bar \nu=\frac12$) which are the two conformal situations.

We are interested in the generation of long-wavelength modes satisfying $\frac{k}{\ke}\ll 1$. In the limit 
\beq
k\ll a~H=\frac{a'}{a}\sim\frac{1}{\eta}\Rightarrow~~ k\eta \ll 1\,,
\eeq
and using
\beq
K_{\bar \nu}(x)~\xrightarrow{x\rightarrow 0}~ \frac{1}{2}\Gamma(|\bar \nu|)\left(\frac{2}{x}\right)^{|\bar \nu|}\,,
\eeq
we obtain for $\bar{\nu}>0$
\bea
&&
    X_k^{(2)}(\eta) \sim k^{-\bar{\nu}} \bar{\eta}^{(\frac{1}{2}-\bar{\nu})}\frac{\mathcal{C}}
{\sqrt{\pi}}
e^{i\left(3\bar \mu \frac{k}{\ke}+\frac{\pi}{4}-\frac{\pi\bar \nu}{2}\right)}
  \label{smallarg}
\\
&& X_k^{(2)'}(\eta)
    \sim k^{-\bar{\nu}}
    \bar{\eta}^{-(\frac{1}{2}+\bar{\nu})}
    \frac{\mathcal{B}}{\sqrt{\pi}}
   e^{i\left(3\bar \mu \frac{k}{\ke}+\frac{\pi}{4}-\frac{\pi \bar \nu}{2}\right)}\,,
  \nonumber
\eea
with 
\begin{align}\label{BC}
    &\mathcal{B}=2^{(\bar{\nu}-1)}\bar{\nu}\Gamma(\bar{\nu})+2^{(\bar{\nu}-2)}\Gamma(\bar{\nu})-2^{\bar{\nu}}\Gamma(\bar{\nu}+1)\nonumber\\\,
&\mathcal{C}=2^{(\bar{\nu}-1)}\Gamma(\bar{\nu})\nonumber\,.
    \end{align}
    
Implementing Eqs.~(\ref{fregion1}) 
and  (\ref{smallarg}) into (\ref{bogo}), we 
can (at last) compute the long-wavelength spectrum 
of massless particles during the transition from the de Sitter phase to the Universe with a generic EoS $w_\phi$
\begin{equation}\label{longspectra} 
\boxed{|\beta_k|^2_{\text{IR}}=\frac{\mathcal{D}}{2\pi}\left(\frac{\ke}{k}\right)^{(2\bar{\nu}+3)}}\,,
\end{equation}
or
\beq
\boxed{
a^3 \left.\frac{dn_\chi}{d \ln k}\right|_{\rm IR}=2\mathcal{D} \frac{\ke^3}{(2 \pi)^3}
\left(\frac{\ke}{k}\right)^{2\bar{\nu}}}
\eeq
with 
\beq
\label{D}
    \mathcal{D}=\left(\mathcal{B} \big(3\bar{\mu}-1\big)^{-\frac12-\bar{\nu}}-\mathcal{C}\big(3\bar{\mu}-1\big)^{\frac12-\bar{\nu}}\right)^2\,
    \eeq
and $\bar \nu$ and $\bar \mu$ given by Eqs.(\ref{Eq:mubar}) and (\ref{Eq:nubar}) respectively. The solution $\beta_k$  has been computed by evaluating Eq.~(\ref{bogo}) 
at the junction point $\eta=\eta_{\rm e}=-\frac{1}{a_{\rm e} H_{\rm e}}=-\frac{1}{\ke}$, or $ |k\eta_{\rm e}|=\frac{k}{\ke}$.
In this IR spectrum for massless fluctuations, the spectral index $(-2\bar \nu -3)$, goes from -6 to -3 for $0 \leq w_{\phi}\lesssim 1$. 
We show in Fig.~(\ref{fullspectrumfig}) the spectrum obtained by solving numerically the set of equations, in comparison with our analytical expression (\ref{longspectra}), for three values of $w_\phi$ (0, $\frac12$ and $\frac23$). In Fig.(\ref{fullspectrumfig}), to numerically track the evolution of the IR modes $k\lesssim \ke$ from the beginning of early inflation to the horizon reentry during reheating, we solve the dynamical equation (\ref{fdynamical2}) along with (\ref{fnoden}) taking the Bunch-Davies vacuum state as an initial state of the field at the beginning of inflation. The non-adiabatic evolution of these large scales after horizon exit during inflation is numerically captured by solving this set of equations, setting the initial time when the largest scale (or the smallest mode) under study was deep inside the horizon during inflation, and the final time is set when this largest scale is well inside the horizon during reheating. We observe a good agreement in the low-frequency regime, 
with a slope in the spectrum $|\beta_k|^2\propto k^{-6}$ for $w_\phi=0$, $\propto k^{-\frac{18}{5}} $ for $w_\phi=\frac12$ and $\propto k^{-\frac{10}{3}}$ for $w_\phi=\frac23$. Note that 
the density spectrum becomes flat $\left(\frac{dn_\chi}{d \ln k}\sim |\beta_k|^2 k^3 \sim \text{constant} \right)$ for kination ($w_\phi\rightarrow 1,n\rightarrow \infty, \bar \nu \rightarrow 0$). 

\subsubsection{Effects of a finite mass term $m_\chi$}

\begin{figure*}
\begin{center}
\includegraphics[width=0.8\linewidth]{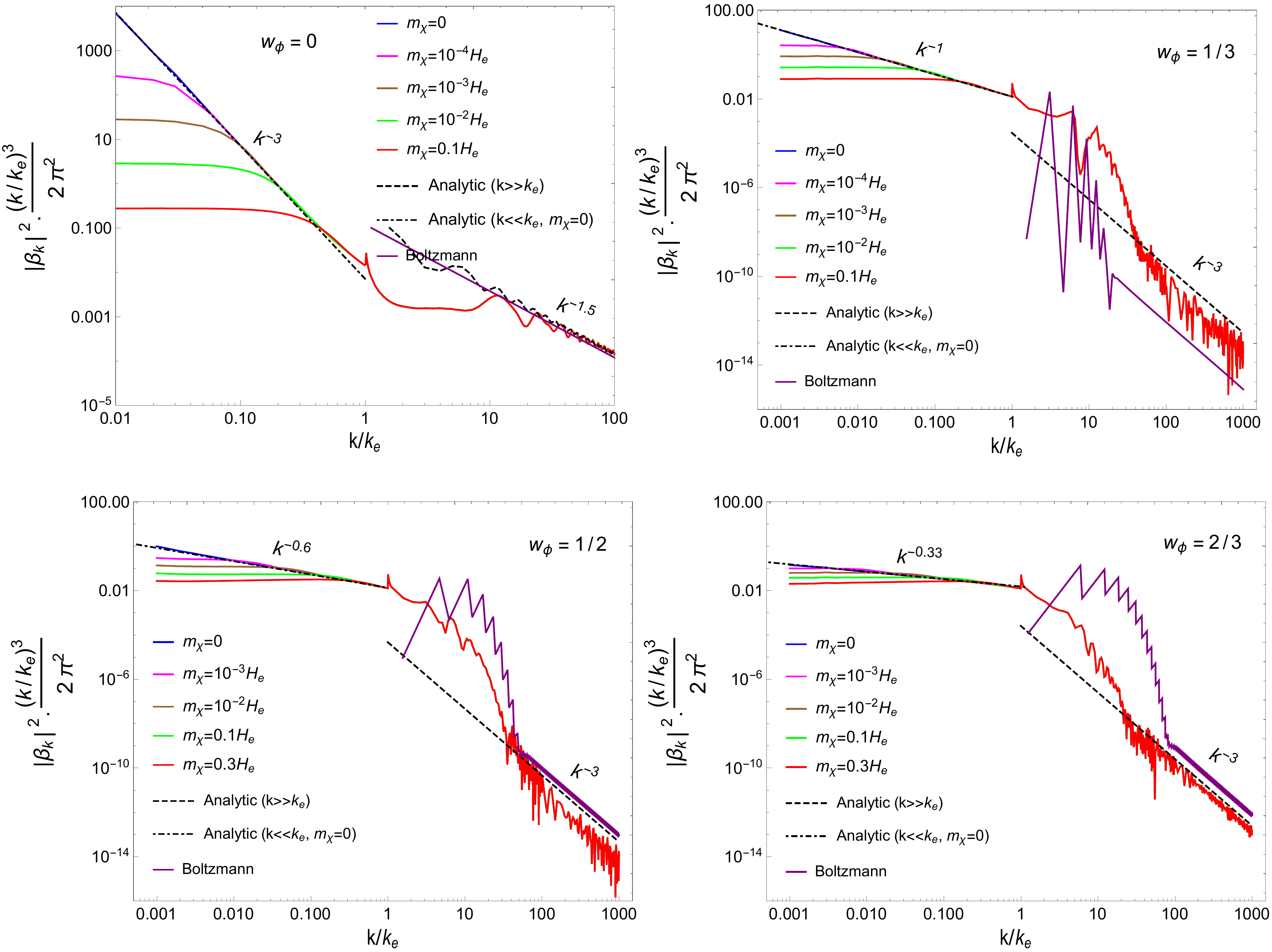}
\caption{\textit{Spectral density $\frac{dn_\chi}{d\log k}$ in long- and short-wavelength regimes, each panel for a different EoS. \textbf{Top-Left:}  $w_{\phi}=0$. \textbf{Top-Right:} $w_{\phi}=1/3$. \textbf{Bottom-Left:} $w_{\phi}=1/2$. \textbf{Bottom-Right:}  $w_{\phi}=2/3$. 
The long-wavelength approximations $k\ll \ke$
from Eq.(\ref{longspectra}) are represented in dash-dotted black for the massless case $m_{\chi}=0$. The short-wavelength approximations 
$k\gg \ke$ from Eqs. (\ref{beta0.3f}) and (\ref{smallspectra}) are represented in dashed black. Numerical results are represented in colored solid lines for different values of $m_\chi$. The short-wavelength spectral density obtained through the Boltzmann treatment in Eqs. (\ref{Boltzmaneqf}) and (\ref{boltzspec0p33}) are depicted in solid purple. For all the EoS, the shape of the short-wavelength $(k\gg k_e)$ perturbative spectral density obtained through a Boltzmann approach is in accord with the non-perturbative Bogoliubov spectra, although a difference in amplitude is observed. As well, the
long-wavelength approximation
Eq.(\ref{longspectra}) is consistent with the numerical result for the massless case, while the short-wavelength approximation Eqs. (\ref{beta0.3f}) and (\ref{smallspectra}) are fully consistent with the numerical result deep in the UV. Interestingly, the interference term in (\ref{beta0.3f}) nicely explains the high-frequency oscillation in the spectrum for $w_\phi = 0$.
}}
\label{fullspectrumfig}
\end{center}
\end{figure*}

Remark that the spectrum is also flat in the case of finite mass, for $k \ll m_\chi$, when 
the equation of motion becomes almost independent of $k$. This behavior was already highlighted in recent works \cite{Garcia:2022vwm, Kaneta:2022gug, Kolb:2023ydq, Jenks:2024fiu} for EoS $w_\phi = 0$, and we extend it here for higher EoS.
To obtain the scale at which the spectrum of long wavelength becomes flat due to the mass term of the mode, we can 
look at Eq.~(\ref{fdynamical3}) in the massive case. We then have the dependence of the frequency $\omega_k^2(\eta)$ as a function of the mass of the mode in Eq.~(\ref{Eq:omegakH}). For convenience in the following discussion let us rewrite 
\begin{align}
\omega_k^2(\eta)=\bigg(k^2+a^2m_{\chi}^2+\left(\frac{3 w_\phi-1}{2}\right)a^2H^2\bigg)
\label{Eq:omegakH2}
\,.
\end{align}
We first determine for which scale factor $a_m$ the mass term $a^2m_\chi^2$ becomes comparable to the expansion term $\propto a^2 H^2$ and find the value of the Hubble scale at this scale factor $H_m \equiv H(a_m)$,
\begin{equation}
\frac{a_m}{a_{\rm e}} = \left(\sqrt{\frac{2}{|3w_\phi -1|}}\frac{m_\chi}{H_{\rm e}}\right)^{-\frac{2}{3(1+\wre)}} \,,
\end{equation}
or
\begin{equation}
H_m = \sqrt{\frac{2}{|3w_\phi -1|}}m_\chi \, .
\end{equation}
Then, it is easy to determine for which scales $k\lesssim k_m$, at horizon reentry after inflation, the mass term dominates over the expansion term, leading to a flat spectrum. We have $k_m \equiv a_m H_m$ and we find the following expression as a function of $w_\phi$ and $m_\chi$
\begin{equation}
\frac{k_m}{\ke} = \left(  \sqrt{\frac{2}{|3w_\phi -1|}}\frac{m_\chi}{H_{\rm e}}\right)^{\frac{1+3w_\phi}{3(1+w_\phi)}}
\end{equation}
which matches perfectly the turning point on the different spectra obtained numerically in Figure \ref{fullspectrumfig}.
Note that for $\wre=1/3$, as the expansion term $\propto a^2H^2$ vanishes in Eq.~(\ref{Eq:omegakH2}), the instant $a_{m}/\ae$ becomes
\begin{equation}\label{am0.3}
\frac{a_m}{\ae}=\sqrt{\frac{\He}{m_{\chi}}} \,,
\end{equation}
which gives
\begin{equation}\label{km0.3}
  \frac{k_m}{\ke}=\sqrt{\frac{m_{\chi}}{\He}}  \,,
\end{equation}
which is also what we observe in Fig.~(\ref{fullspectrumfig}).
So, any $k\lesssim k_m$ suffers from the finite mass-breaking effect of the IR spectrum.

To get further insight, it is interesting to remark 
that for $\wre=1/3$, there exists an exact solution for the mode 
equation of the massive field, that we propose to compute.  
In terms of Hankel function, the inflationary solution of the massive field is given by
\begin{equation}\label{infmass}
  X_k^{(1)}(\eta)=\frac{\sqrt{-\pi \eta}}{2}e^{i(\pi/4+\pi\bar{\nu}_1/2)}H^{(1)}_{\bar{\nu}_1}(k|\eta|)  \,,
\end{equation}
where the (mass-dependent) index is now 
\beq
\bar{\nu}_1=\sqrt{\frac{9}{4}-\frac{m_{\chi}^2}{\He^2}}\,.
\eeq
The adiabatic out-vacuum solution of (\ref{fdynamical3}) during reheating for finite mass, $(\frac{m_{\chi}}{\He})<3/2$ and $\wre=1/3$ is
\begin{equation}\label{rehmass0p33}
 X_k^{(2)}(\eta)=\frac{e^{-\frac{\pi k^2}{8 m_{\chi}\ae^2\He}}}{(2 m_{\chi}\ae\ke)^{1/4}}D_{\bar{\nu}_2}\Big(e^{i\frac{\pi}{4}}\sqrt{\frac{2 m_{\chi}}{\He}}(\eta\ke+2)\Big)\times e^{i\delta}\,,  \end{equation}
where $D_{\bar{\nu}_2}$ is Parabolic Cylinder function with $\bar{\nu}_2=-\frac{1}{2}\big(1+\frac{i k^2}{m_{\chi}\ae^2\He}\big)$, and the phase 
\beq
\delta=\frac{\pi}{8}-\frac{ k^2}{4 m_{\chi}\ae^2\He}\text{ln}(2)-\frac{3 k^2}{4 m_{\chi}\ae^2\He}\ln{\left(\frac{m_{\chi}}{\He}\right)}
\,.
\eeq 
Using (\ref{infmass}) and (\ref{rehmass0p33}) in (\ref{bogo}) we obtain the spectral 
nature in the long-wavelength regime as 
\beq
\boxed{
|\beta_k|_{{\rm IR},m_\chi \neq 0}^2\sim \frac{e^{-\frac{\pi \He }{4 m_{\chi}}\left(\frac{k}{\ke}\right)^2}}{\sqrt{m_{\chi}/\He}}\left(\frac{\ke}{k}\right)^{2\bar{\nu}_1}\,.}
\label{Eq:betakir}
\eeq
In the regime $(k/\ke)\ll\sqrt{(m_{\chi}/\He)}$ and $(m_{\chi}/\He)\ll3/2$, the spectral nature turns out 
to be $|\beta_k|^2\propto \big(\ke/k\big)^{3}$, which justifies 
the behavior of the spectra for finite mass in the top right 
panel of Fig.(\ref{fullspectrumfig}). This result also agrees with the condition (\ref{km0.3}). 

As the general mode solution 
doesn't exist for any higher Eos $\wre>1/3$, we resort to the 
approximated method as outlined in \cite{Kaneta:2022gug} to 
determine at least the behavior of the number density spectrum for the finite mass case.  We write the WKB approximated post-inflationary solution as
\begin{equation}\label{rehmassw}
   X_k^{(2)}(\eta)\simeq \frac{e^{-i \Omega_k(\eta)}}{\sqrt{2\omega_k(\eta)}}
\end{equation}
Using the equations (\ref{infmass}) and (\ref{rehmassw}) in (\ref{bogo}) we find
\begin{align}\label{betamassw}
\beta_k\simeq & \sqrt{\frac{\pi}{32 \tilde{\omega}_k(\etae)}} \bigg[\left(\frac{k}{\ke}\right)\left(H^{(1)}_{\bar{\nu}_1+1}(k/\ke)-H^{(1)}_{\bar{\nu}_1-1}(k/\ke)\right)\nonumber\\
&+\left(\frac{\left(\frac{d\tilde{\omega}_k(\etae)}{d(\eta\ke)}\right)}{\tilde{\omega}_k(\etae)}-1+2 i\tilde{\omega}_k(\etae)\right)H^{(1)}_{\bar{\nu}_1}(k/\ke)\bigg]\times e^{i \delta}
\end{align}
Here the phase $\delta~=~i\left(\frac{3\pi}{4}+\frac{\pi\bar{\nu}_1}{2}-\Omega_k(\eta)\right)$.
At the end of inflation ($\eta=\etae$), the dimensionless terms are given by
\begin{align}\label{dimlessomega}
  &\tilde{\omega}_k(\etae)=\sqrt{(k/\ke)^2+(m_{\chi}/\He)^2+(3\wre-1)/2}\, ,\nonumber\\
  &\frac{\left(\frac{d\tilde{\omega}_k(\etae)}{d(\eta\ke)}\right)}{\tilde{\omega}_k(\etae)}=\left(\frac{\He}{m_{\chi}}\right)^{5/2}\left(1+\left(\frac{\He}{m_{\chi}}\right)^2\frac{(3\wre-1)}{2}\right)^{-9/4}
  \end{align}
Likewise $\wre=1/3$, in the long-wavelength limit $k/\ke\ll1$, Eq.~(\ref{betamassw}) gives the spectral nature $|\beta_k|^2\propto \left(\ke/k\right)^3$ for the masses $\left(m_{\chi}/\He\right)\ll3/2$, which is indeed independent of EoS. For the mass $\left(m_{\chi}/\He\right)>3/2$, the mass-dependent index $\bar{\nu}_1$ becomes imaginary, this in turn generates an exponential mass-suppressed amplitude of the IR spectrum from the phase part $e^{i \delta}$ in the Eq.(\ref{betamassw}).  \\

Before closing this section, we notice that for light scalar modes with finite mass term $(m_\chi/H_e)\ll 3/2$, in the long-wavelength limit $k/k_e\ll 1$, the analytic calculations provide flat spectra, whatever the EoS and the specific shape of the inflaton potential. However, it has been shown that the numerical spectra in this limit can be slightly red-tilted depending on the specific inflationary model and evolution of the Hubble scale during inflation \cite{Jenks:2024fiu}. Our analytic approximation assumes an initially constant Hubble rate during the de Sitter phase, which produces a flat long-wavelength spectrum. This is a reasonable assumption for a wide range of inflationary models, corresponding to an inflaton that begins in a very flat potential region, which is valid for $\alpha$-attractor models that we consider as benchmarks in this work. We recover in our numerical results flat spectra in the long-wavelength limit, as can be seen in Figure \ref{fullspectrumfig} for such inflationary potentials.
\subsection{Sub-horizon modes spectrum ($k>a_{\rm e} H_{\rm e}$)}

\subsubsection{Generalities}

For modes staying inside the horizon during the whole inflationary phase(never experienced adiabatic to non-adiabatic transition), particle production essentially takes 
place almost immediately after the end of inflation within a very few e-folding 
numbers, and the coherently oscillating inflaton background becomes the source of the production.
Typically, the produced particle number density $|\beta_k|^2$ is 
bound to be very small, because the divergent term 
$\propto \frac{1}{k \eta}$ is not present in the solution 
$X_k^{(1)}$ in Eq.~(\ref{fregion1}), $\omega_k^2$ being always positive.
However, the equation of state varying from $-1$ to $w_\phi$,
one still has the gravitational production of a certain amount
of modes.

In this case, the solution stays adiabatic during all 
the production process, and one does not need to compute 
the exact solutions $X_k^{(1)}(\eta)$ and $X_k^{(2)}(\eta)$ to find the late-time converging quantity $\beta_k$. Instead, one can compute it directly from the evolution equation of the time-dependent coefficient. Indeed, implementing Eq.~(\ref{WKBform})
into
\beq
X_k'' + \omega_k^2 X_k=0\,,
\eeq
gives in the adiabatic approximation (neglecting second-order derivatives, and considering $\alpha_k \gg \beta_k$)
\cite{Kofman:1997yn,Ema:2015dka,Ema:2016hlw,Chung:2018ayg,Basso:2021whd,Ema:2018ucl},
\beq
\beta_k'(\eta)=\frac{\omega_k'}{2 \omega_k}e^{-2 i \Omega_k(\eta)}\,,
\eeq
or
\begin{equation}\label{Eq:mybeta1}
   \beta_k(\eta)\simeq \int^{\eta}_{\eta_{\text{e}}} d\eta^{'} \frac{\omega_k^{\prime}}{2\omega_k}e^{-2i\Omega_k(\eta^{'})}\,,
\end{equation}
where 
\beq
\Omega_k(\eta^{'})=\int_{\eta_{\text{e}}}^{\eta^{'}}\omega_k(\eta)d\eta=\int_{t_{\text{e}}}^{t^{'}}\frac{\omega_k(t)}{a(t)}dt
\eeq
and $t_{\text{e}}$ stands for the time at the end of inflation.
This approach of computing the Bogoliubov coefficient has the advantage of finding the time-independent $\beta_k$ as the limit $\beta_k(\eta\rightarrow \infty)$, without the need to find the explicit solution $X_k^{(2)}(\eta)$.

Using the time-dependent frequency expressed in Eq.~(\ref{Eq:omegakH}), and working in cosmic 
time coordinate, we can express
\beq
a'=a^2H~\Rightarrow~\frac{a''}{a} =2 a' H + a H'=a^2(2H^2+\dot H)\,,
\label{Eqgravsource}
\eeq
which implies that Eq.(\ref{Eq:omegakH}) can be written
\beq
\omega^2_k(t)=k^2+a^2(m_\chi^2-2H^2-\dot H)\,,
\eeq
or
\beq
\frac{\dot \omega_k(t)}{\omega_k(t)}=\frac{a^2}{\omega_k^2}
\left[Hm_\chi^2-2H^3-3H\dot H-\frac12\ddot H\right]\,.
\label{Eq:mydot}
\eeq

Using the development (\ref{Hubbleforn}) for $H$, and
with the help of Eq.~(\ref{Pdynamical}),
we get at the leading order in $\Phi(t)/M_{P}$,
the behavior of $\dot H(t)$ and $\ddot H(t)$ for general $n$
\beq
\dot H(t) 
\simeq   3\bar{H}^2\bigg[\left(\mathcal{P}^{2n}-1\right)-\frac{\sqrt{6}\mathcal{P}\sqrt{1-\mathcal{P}^{2n}}}{(n+1)}\left(\frac{\Phi(t)}{M_{P}}\right)\bigg]\,,
\label{Eq:hdot}
\eeq
and
\bea
  \ddot{H}(t) \simeq && \frac{9 \bar H^3}{(n+1)}\bigg[\Big(4-(4+2n){\mathcal P}^{2n}\Big)\nonumber\\ &&+\left(\frac{\Phi(t)}{M_{P}}\right)\sqrt{6}\mathcal{P}\sqrt{1-\mathcal{P}^{2n}}\Big(\frac{n+4}{n+1}-\mathcal{P}^{2n}\Big)\bigg]
  \nonumber\\
    \nonumber
    \\
    && -6\sqrt{6}n\bar{H}^3 \left(\frac{M_{P}}{\Phi(t)}\right)\mathcal{P}^{2n-1}\sqrt{1-\mathcal{P}^{2n}}
    \,.
\eea
From Eq.~(\ref{Hubbleforn}), the leading terms of $H^3$ can be written as
\beq
\label{Hcube}
H^3(t)\simeq \bar{H}^3\bigg(1+\frac{3\sqrt{6}\mathcal{P}\sqrt{1-\mathcal{P}^{2n}}}{2(n+1)}\left(\frac{\Phi(t)}{M_{P}}\right)\bigg)\,.
\eeq
Finally, using the expressions of $H(t)$ and $\dot H(t)$ in Eqs.~(\ref{Hubbleforn}) and (\ref{Eq:hdot}) and keeping only the terms up to the first order of ($\Phi(t)/M_{P}$) we can write
\bea
\label{HHdot}
&&
 H(t)\dot H(t) \simeq 3\bar{H}^3\bigg[
 \Big(\mathcal{P}^{2n}-1\Big)
 \nonumber
 \\
&&
+\frac{\sqrt{6}\mathcal{P}\sqrt{1-\mathcal{P}^{2n}}(\mathcal{P}^{2n}-3)}{2(n+1)}\left(\frac{\Phi(t)}{M_{P}}\right)\bigg]\,.
\eea
Implementing these developments in Eq.~(\ref{Eq:mydot}), we then obtain for $k \gg a H$,

\bea
  \frac{\dot{\omega}_k}{\omega_k}\simeq 
  &&
  \frac{1}{\left(\frac{k^2}{a^2}+ m^2_{\chi}\right)}
  \bigg[H m^2_{\chi}+
  \left(\frac{7n-11}{n+1} +\frac{9}{n+1}{\cal P}^{2n}
  \right)\bar H^3
  \nonumber
  \\
  && +\frac{3\sqrt{6}\bar H^3\mathcal{P}\sqrt{1-\mathcal{P}^{2n}}}{2(n+1)^2}(4n-5)\left(\frac{\Phi(t)}{M_{P}}\right)\nonumber\\
   && +3\sqrt{6}n \bar H^3\left(\frac{M_{P}}{\Phi(t)}\right)\mathcal{P}^{2n-1}\sqrt{1-\mathcal{P}^{2n}}
 \bigg]  \,.
    \label{Eq:mydotomegak2}
\eea

\noindent
Writing $\bar H = \He \left( \frac{\te}{t}\right)$ and 
$\Phi(t)=\phi_e\left(\frac{\te}{t}\right)^\frac{1}{n}$, 
we can substitute (\ref{Eq:mydotomegak2}) back to Eq.(\ref{Eq:mybeta1}),
with $d\eta' \frac{\omega_k'}{\omega_k}=dt'\frac{\dot \omega_k}{\omega_k}$.
$\beta_k$ then becomes
\begin{multline}\label{Eq:mybeta4}
 \beta_k \simeq 
 \frac{1}{2}\sum_{\nu,l\neq 0}\int_{t_{\text{e}}}^{t}dt^{'}
 \left(\frac{\te}{t'}\right)^3
 \Bigg[ {\cal N}_0e^{i(\nu+l)\omega t^{'}} 
 \left(\frac{t'}{\te}\right)^\frac{1}{n}
 \\+ {\cal N}_1 e^{i\nu\omega t^{'}}+{\cal N}_2 +~ \mathcal{N}_3e^{i(\nu+l)\omega t^{'}}\left(\frac{\te}{t^{'}}\right)^\frac{1}{n}
 +{\cal N}_4 \left(\frac{t'}{\te}\right)^2\\
 +{\cal N}_5e^{i(\nu+l)\omega t^{'}} \left(\frac{t'}{\te}\right)^{\frac{2n-1}{n}}
\Bigg]
\times \frac{e^{-2i\Omega_k(t^{'})}}{{\left(\frac{k^2}{a^2}+m^2_{\chi}\right)}} \,,
\end{multline}
where we have introduced
\bea
&&
 {\cal N}_0 = 3{\sqrt{6}n {H^3_{\rm e}}\big(\mathcal{P}^{2n-1}\big)_{\nu}\left(\sqrt{1-\mathcal{P}^{2n}}\right)_{l}} \left(\frac{M_{P}}{\phi_{\text{e}}}\right)\,;
 \nonumber
 \\
 && {\cal N}_1=\frac{9}{n+1} 
 \He^3 ~\mathcal{P}^{2n}_{\nu}\,;\quad
 {\cal N}_2=\frac{7n-11}{n+1} \He^3\,;\quad\nonumber\\
 && \mathcal{N}_3=\frac{3\sqrt{6}H_{\rm e}^3\mathcal{P}_{\nu}\left(\sqrt{1-\mathcal{P}^{2n}}\right)_{l}}{2(n+1)^2}\big(4n-5\big)\left(\frac{\phi_{\rm e}}{M_{P}}\right)\,;\quad\nonumber\\ 
  && {\cal N}_4=m_{\chi}^2 H_{\rm e}\,\nonumber\\
  && {\cal N}_5={\cal N}_4 \frac{\sqrt{6}\mathcal{P}_{\nu}\left(\sqrt{1-\mathcal{P}^{2n}}\right)_{l}}{2(n+1)}\left(\frac{\phi_{\rm e}}{M_{P}}\right)\,.
 \label{Eq:ni}
\eea
\noindent
Although it is justified to keep only up to the first order terms 
of $\left(\frac{\Phi(t)}{M_{P}}\right)$ for the sake of comparison between the nature of 
the spectra in the UV regime for two approaches 
(Boltzmann and Bogoliubov), to achieve an amplitude-wise 
uniformity between the analytical and the numerically obtained spectrum, it becomes imperative 
to consider the higher order terms of $\Big(\frac{\Phi(t)}{M_{P}}\Big)$. Although these higher order terms, decaying 
rapidly with time, hardly contribute at any late time, from the perspective of initial time $t_{\rm e}$, they have a non-
negligible contribution in the total amplitude. This amplitude will be also important for a correct estimate of reheating temperature using this spectrum later. Hence, it is important to 
keep them while computing the integral (\ref{Eq:mybeta1}). A complete form of the $\beta_k$ spectrum in the UV regime including the higher order terms is given in the Appendix \ref{appenA}. 
Here we shall proceed with the integral (\ref{Eq:mybeta4}) with the first-order term to compute the nature of the UV spectrum. 

In Eq.~(\ref{Eq:ni})
we have introduced separate Fourier series 
having Fourier components $\mathcal{P}_{\nu}, \left(\sqrt{1-\mathcal{P}^{2n}}\right)_{\nu}$ and $ \big(\mathcal{P}^{2n-1}\big)_{\nu}$. In 
Eq.~(\ref{Eq:mybeta4}), there are terms involving the product of oscillatory phase 
part and non-oscillatory decaying part, as well as non-oscillatory decaying part only. 
Due to their specific nature, each part needs to be considered with a separate analytic treatment.
To evaluate the integral associated with a term involving the product of 
oscillatory part and non-oscillatory part (like the first term $\mathcal{N}_0$), we shall use stationary 
phase approximation \cite{Ablowitz_Fokas_2003, Rozman2017, Kolb:2023ydq} and find out the stationary point of the total phase part, $\varphi_{\pm}(t)\equiv (\nu+l)\omega t-2\Omega_k(t)$ ($\nu,l$ may be both 
positive and negative).
Defining the time instant of the stationary phase to be $t_k$ at which $\dot{\varphi}_{\pm}(t_k)=0$, it is 
trivial to check that $\varphi_{-}$ does not have any solution to satisfy the 
condition $\dot{\varphi}_{-}(t_k)=0$, $\Omega_k(t)$ being always positive. Therefore, from now on, we shall only work 
with $\varphi_{+}(t)$ to determine the stationary phase solution. The mass term $m_{\phi}$ appearing in the fundamental frequency $\omega$ of the background inflaton, Eq.~(\ref{fundafre}), is a function of time for $n\neq 1$. Considering this time-dependent mass term, 
we get the following constraint relation of stationary point
 \begin{align}\label{stationaryp}
     &\dot{\varphi}_{+}(t_k)=0   \Rightarrow (\nu+l)\Big(\omega|_{t=t_k}+\dot\omega|_{t=t_k}t_k\Big)=\frac{2(\omega_k|_{t=t_k})}{a_k}\nonumber\\
    &\Rightarrow \left(\frac{(\nu+l)\bar{\alpha}m^{\rm e}_{\phi}}{2n}\right)^2\left(\frac{a_k}{a_{\rm e}}\right)^{\frac{8-4n}{n+1}}\simeq\left(\frac{k^2}{a_{\rm e}^2}+\left(\frac{a_k}{a_{\rm e}}\right)^2m_{\chi}^2\right)
 \end{align}
 where 
 \beq
\bar{\alpha}=\sqrt{\frac{\pi n}{(2n-1)}}\frac{\Gamma\left(\frac{1}{2}+\frac{1}{2n}\right)}{\Gamma\left(\frac{1}{2n}\right)}\,,
\label{Eq:alphabar}
 \eeq
 $a_k\equiv a(t=t_k)$, 
 and we supposed $\ae \ll a_k$ in the right--hand side of the equation. The time 
 instant $t_k$ should be within the range $t_{\text{e}}<t_k<t$ otherwise, the integrand is highly oscillating and that 
 gives vanishing contribution. For the purpose of reheating, the produced 
 particles can be assumed to be\ massless or $\frac{m_{\chi}}{m_{\phi}^{\rm e}}\ll1$ 
 such that ${\cal N}_3, {\cal N}_4 \approx 0$. However, for dark matter, those terms 
 may have a significant contribution. Nonetheless, in the following discussions 
 we consider this approximation, and for such case one gets the exact solution for the above stationary phase equation as 
 \begin{align} \label{stationarysol}
   \frac{a_k}{a_{\rm e}} \propto 
   k^{\frac{n+1}{4-2n}}=
     k^{\frac {1}{1-3w_{\phi}}} 
   \end{align}
  In the high-frequency limit, for $w_{\phi} < 1/3$, it is indeed clear that $a_{\rm e} < a_k < a(t)$, and we can clearly have a stationary point within the integration range. On the other hand for $w_{\phi} > 1/3$, one obtains $a_k < a_{\rm e} < a(t)$ which remains always outside the range of integration, and does not have any stationary point. Therefore, in the following, we compute the number spectrum for two different regimes of equation of states separately. 
  
\subsubsection{Spectrum for EoS \quad $0\leq w_{\phi}<1/3$}

 In the  large $k$ limit, considering the dominant term from (\ref{Eq:mybeta4}), we, therefore, write the spectrum for this EoS range as
 \begin{multline}\label{beta0.31}
    \beta_k\simeq \frac{1}{2}\sum_{\nu,l\neq 0}\int_{t_{\text{e}}}^{t}dt^{'}\Bigg[ {\cal N}_0e^{i(\nu+l)\omega t^{'}} \left(\frac{a(t^{'})}{a_{\text{e}}}\right)^{-3(1+2w_{\phi})}\\+ {\cal N}_2 \left(\frac{a(t^{'})}{a_{\text{e}}}\right)^{-\frac{9(1+w_{\phi})}{2}}\Bigg]
\times \frac{e^{-2i\Omega_k(t^{'})}}{{\big(k^2/a^2\big)}} 
 \end{multline}

One clearly sees that the first term in the 
above integral has stationary points for $0\leq w_{\phi}<1/3$, 
whereas the second term does not have such points. Therefore, 
for the first integral one can readily use the stationary phase 
approximation whereas, the leading contribution for large $k$ 
from the second term can be obtained by utilizing the simple 
integration by parts method.\\ 

Employing the stationary phase approximation method, we write the approximate form of the integral (\ref{beta0.31}) associated with ${\cal N}_0$ term as 
\begin{equation}\label{beta0.32}
   \beta_k^{(0)}\simeq \sum_{\nu,l>0}\frac{\mathcal{N}_0 a^2_{\rm e}}{2k^2} \left(\frac{a_k}{a_{\rm e}}\right)^{-(1+6w_{\phi})}\sqrt{\frac{2\pi}{|\ddot\varphi_{+}(t_k)|}}e^{\pm i\frac{\pi}{4}+i\varphi_{+}(t_k)} 
\end{equation}
Where double derivative of total phase part $\ddot\varphi_{+}(t_k)$ is calculated to be
\begin{equation}\label{ddotphase0.3}
  \ddot\varphi_{+}(t_k)=2(1-3w_{\phi})H_{\rm e}\bar{\beta}\left(\frac{\bar{\beta} a_{\rm e}}{k}\right)^{\frac{3(1+3w_{\phi})}{2(1-3w_{\phi})}} \,,
\end{equation}
with $\bar{\beta}=\left(\frac{(\nu+l)(1-w_{\phi})\bar{\alpha}m^{\rm e}_{\phi}}{2(1+w_{\phi})}\right)$.
Implementing (\ref{ddotphase0.3}) in (\ref{beta0.32}), we obtain
\begin{align}\label{beta0.33}
   \beta_k^{(0)}\simeq & \sum_{\nu,l>0}\frac{\mathcal{N}_0\bar{\beta}^{p_1}}{2H_{\rm e}^{p_2}}\sqrt{\frac{\pi}{|(1-3w_{\phi})|}}\left(\frac{a_{\rm e}H_{\rm e}}{k}\right)^{\frac{9(1-w_{\phi})}{4(1-3w_{\phi})}}e^{\pm i\frac{\pi}{4}+i\varphi_{+}(t_k)} \nonumber\\
  & = \bar{\cal N}_0{k}^{\frac{9(w_{\phi}-1)}{4(1-3w_{\phi})}}e^{\pm i\frac{\pi}{4}+i\varphi_{+}(t_k)}\,,
\end{align}
where $\bar{\cal N}_0=\sum_{\nu,l>0}\frac{\mathcal{N}_0\bar{\beta}^{p_1}}{2H_{\rm e}^{p_2}}\sqrt{\frac{\pi}{|(1-3w_{\phi})|}}\Big(a_{\rm e}H_{\rm e}\Big)^{\frac{9(1-w_{\phi})}{4(1-3w_{\phi})}}$,
 $p_1=\frac{21w_{\phi}-1}{4(1-3w_{\phi})}$ and $p_2=\frac{11-15w_{\phi}}{4(1-3w_{\phi})}$.\\
 
 In the integral (\ref{Eq:mybeta4}), the oscillatory terms associated with the coefficients $\mathcal{N}_1$ and $\mathcal{N}_3$ also have stationary points. Following exactly the same approach as is followed to reach the expression (\ref{beta0.33}) by evaluating the integral (\ref{beta0.31}) for $\mathcal{N}_0$, we find the spectra $\propto k^{\frac{15( w_{\phi}-1)}{4(1-3w_{\phi})}}$ and $\propto k^{\frac{21(w_{\phi}-1)}{4(1-3w_{\phi})}}$ for the coefficients $\mathcal{N}_1$ and $\mathcal{N}_3$ respectively. Evidently, the presence of a faster time-decaying term in the amplitude of the coefficient $\mathcal{N}_3$ in Eq.(\ref{Eq:mybeta4}) essentially causes a sub-dominant spectral index in the associated number density spectrum as compared to $\mathcal{N}_1$ for any $0\leq \wre<1/3$. Additionally, both the indices for $\mathcal{N}_1$ and $\mathcal{N}_3$ are found to be subdominant as compared to the index ${k}^{\frac{9(w_{\phi}-1)}{4(1-3w_{\phi})}}$ associated with the coefficient $\mathcal{N}_0$ for any $w_{\phi}<1/3$ in the large frequency limit. Hence, the oscillatory terms with the coefficients $\mathcal{N}_1$ and $\mathcal{N}_3$ have been neglected in the integral (\ref{beta0.31}).\\  
 
As mentioned earlier, the second part of the integral 
(\ref{beta0.31}) $\propto {\cal N}_2$,
which does not contain an oscillatory part, 
 can be solved by an integration by parts.
 For that, one needs to consider the generic form of integral
 \begin{align}\label{Iint}
I(x)&=\int_{a}^{b}g(t)e^{ix\psi(t)}dt\nonumber\\&=\frac{1}{ix}\int_{a}^{b}\frac{g(t)}{\dot\psi(t)}d(e^{ix\psi(t)})\nonumber\\ 
&= \frac{1}{ix}\frac{g(t)}{\dot\psi(t)}e^{ix\psi(t)}\Big|_{a}^{b}-\frac{1}{ix}\int_{a}^{b}\frac{d}{dt}\left(\frac{g(t)}{\dot\psi(t)}\right)e^{ix\psi(t)}dt\,.
 \end{align}
 Because $\psi(t)$ has no 
 stationary point in the given range $[a,b]$, i.e $\dot\psi(t)\neq 0$, a simple integration by parts gives a leading 
 asymptotic behavior (Riemann–Lebesgue lemma) \cite{Ablowitz_Fokas_2003, Rozman2017} 
 \begin{equation}\label{Iasymp}
     I(x)\sim \frac{1}{ix}\frac{g(t)}{\dot\psi(t)}e^{ix\psi(t)}\Big|_{a}^{b}\quad \quad \text{as}\quad x\rightarrow \infty\,.
 \end{equation}
  Comparing the integral (\ref{beta0.31}) with (\ref{Iasymp}) we have for $k\gg a_{\rm e}H_{\rm e}$, 
 \bea
 \label{xpsi}
     && x\equiv k/a_{\rm e}H_{\rm e},\quad \psi(t)\approx -\frac{2H_{\rm e}t}{(a/a_{\rm e})}\,,\nonumber\\
    && \dot\psi(t)\approx \frac{2H_{\rm e}}{(a/a_{\rm e})}\times
\underbrace{
    \left(-\frac{(1+3\wre)}{3(1+\wre)}\right)}_{f(w_{\phi})} \,,\label{Eq:f}\\
    && g(t)= \frac{{\cal N}_2 }{\left(k^2/a^2(t)\right)}\left(\frac{a(t)}{a_{\text{e}}}\right)^{-\frac{9(1+w_{\phi})}{2}}
    \nonumber
 \eea
Finally, the  second part $\propto {\cal N}_2$ of the integral (\ref{beta0.31}) becomes  
\begin{align}\label{beta0.34}
\beta_k^{(2)} & \simeq 
\frac{ia_{\rm e}^3}{4k^3f(w_{\phi})} {\cal N}_2 e^{-\frac{2ikt_{\rm e}}{a_{\rm e}}}\nonumber \\
 &= \bar{\cal N}_2 k^{-3} e^{i\left(\frac{\pi}{2}-\frac{2kt_{\rm e}}{a_{\rm e}}\right)} 
 \end{align}
where $\bar{\cal N}_2=\frac{a_{\rm e}^3}{4f(w_{\phi})}{\cal N}_2$. In the presence of fast decaying terms in the amplitude of Eq.~(\ref{beta0.34}), the dominant contribution comes from the initial time. Note also that, contrary to the 
 oscillatory term $\propto {\cal N}_0$, the spectrum for 
 $|\beta_k^{(2)}|$ {\it does not} depend on the EoS, leading
 to $|\beta_k|^2\propto k^{-6}$.
 
Combining Eq.~(\ref{beta0.33}) and (\ref{beta0.34}), we can then obtain the number density spectrum as $n_k = |\beta_k|^2 = |\beta_k^{(0)} + \beta_k^{(2)}|^2$.
Upon close inspection, one can notice that for the EoS in the range $0\leq w_{\phi}< 1/9$, $|\beta_k^{(0)}|^2$ having a stationary phase contribution dominates over $|\beta_k^{(2)}|^2$ having a non-oscillatory contribution in the high-frequency limit. On the other hand for $ 1/9 <w_{\phi} < 1/3$, it is $|\beta_k^{(2)}|^2$ which dominates, $|\beta_k^{(0)}|^2$ having a steeper spectrum. We emphasize that although in our analytical computation we encounter a critical value $w_\phi = 1/9$, our analysis is performed in the inflaton potential $V(\phi) \propto \phi^{2n}$ with $n$ a positive integer, and EoS is defined as $\wre=(n-1)/(n+1)$. Hence, the value $w_\phi = 1/9$ is not achieved in the class of models of inflation and reheating considered in this work. Still, this is not a limitation of the theoretical framework constructed in this work. We expect such analytical results to hold in any potential allowing for inflaton oscillations near the minimum, as long as we can expand the oscillating background in terms of its Fourier modes.
As a conclusion, the spectral nature of the number density spectrum in the range $0\leq w_{\phi}< 1/3$ can be written as follows
\begin{widetext}
\begin{align}\label{beta0.3f}
\boxed{
  | \beta_k|_{{\rm UV},w_\phi<\frac13}^2=
   \begin{cases}
\left(\bar{\cal N}_0\right)^2 k^{\frac{9(w_{\phi}-1)}{2-6w_{\phi}}}+\underbrace{\bar{\cal N}_0 \bar{\cal N}_2 k^{\frac{45w_{\phi}-21}{4(1-3w_{\phi})}} \cos \psi}_{\text{interference term}} ~~ w_{\phi}\leq 1/9 &\\
\left(\bar{\cal N}_2\right)^2k^{-6}+\underbrace{ \bar{\cal N}_0 \bar{\cal N}_2 k^{\frac{45w_{\phi}-21}{4(1-3w_{\phi})}} \cos \psi}_{\text{interference term}}~~~~~~~ w_{\phi}>1/9 &
\end{cases}
}
\end{align}
\end{widetext}
where the \textit{quantum interference term (product of an oscillatory and a non-oscillatory function) appears as a fast varying 
oscillatory term in the number spectrum with momentum-dependent phase factor}, $\psi=\left({\frac{\pi}{4}+\varphi_{+}(t_k)-2kt_{\rm e}}/a_{\rm e}\right)$. Though the $k$ index of the 
second term in the Eq.~(\ref{beta0.3f}) is sub-leading compared 
to the first one in both cases, it is this term that results in 
oscillation in the spectrum as can indeed be observed in the 
Top left panel of Fig.~\ref{fullspectrumfig}. For $w_{\phi}=0$, we 
indeed recover the leading order spectral behavior $|\beta_k|^2\propto k^{-\frac{9}{2}}$ discussed in \cite{Kaneta:2022gug, Basso:2022tpd}, up to the oscillatory term. With the increase of EoS, this oscillation effect in number density spectrum starts 
getting washed away because of the gradual decay of amplitude of 
the interference term in large $k$ limit, which is evident from the interference term in Eq.~(\ref{beta0.3f}).\\

\subsubsection{Spectrum for EoS \quad  $w_{\phi}\geq1/3$ }

In this section, we further generalize the spectrum for EoS $w_{\phi}\geq 1/3$. The situation becomes drastically different 
in this case. 
Indeed, we have shown that Eq.~(\ref{stationaryp}) has no stationary point within the 
integration range for $\wre\geq 1/3$. As there is no stationary phase in 
this range $w_{\phi}\geq 1/3$, unlike the spectrum (\ref{beta0.31}), we need to take into account 
the contribution of all the terms 
in Eq.~(\ref{Eq:mybeta4}). Therefore, to estimate the 
approximate spectral behavior in this EoS range, we should 
follow the integration by parts method described in the previous section.

Comparing the integral (\ref{Eq:mybeta4}) with (\ref{Iasymp}), for $k \gg a_{\rm e}H_{\rm e}$, in a similar manner we obtain the functions as given in Eq.(\ref{xpsi}).

 We finally end up having the following asymptotic form of the integral (\ref{Eq:mybeta4}) in large $k$ and massless ($m_{\chi}\approx 0$) limit for $w_{\phi}\geq 1/3$.
 \begin{align}\label{beta5}
     \beta_k\simeq& \frac{1}{4if(w_{\phi})(k/a_{\rm e})^3}\sum_{\nu,l\neq 0}\Bigg[ {\cal N}_0 \left(\frac{a(t)}{a_{\text{e}}}\right)^{-6w_{\phi}}\nonumber\\
     &+ \Big({\cal N}_1 +{\cal N}_2\Big) \left(\frac{a(t)}{a_{\text{e}}}\right)^{-\frac{3(1+3w_{\phi})}{2}}+{\cal N}_3 \left(\frac{a(t)}{a_{\text{e}}}\right)^{-3(1+w_{\phi})}\Bigg]\nonumber\\ 
     &\times e^{-\frac{2i(k/a_{\rm e})t}{(a/a_{\rm e})}}\Big|_{t_{\rm e}}^t \,.
 \end{align}
As all the terms in the amplitude part decay fast with time, the dominant contribution will come from the initial time. The final form of the spectrum then becomes 
\begin{widetext}
\beq
\label{smallspectra}
\boxed{
|\beta_k|^2_{{\rm UV},w_\phi \geq \frac13}\simeq \frac{1}{16f^2(w_{\phi})}\left(\frac{a_e}{k} \right)^6\times \sum\sum\Bigg[ {\cal N}_0 + {\cal N}_1 +{\cal N}_2+{\cal N}_3\Bigg]^2\, }
\eeq
\end{widetext}
where $f(w_\phi)$ is given by Eq.~(\ref{Eq:f}).
 This equation represents an approximate spectral behavior in the  UV regime for $w_\phi \geq 1/3$, taking the background inflaton oscillation effect into account. 
 From Eqs.~(\ref{beta0.3f}) and (\ref{smallspectra}), we remark
 that the spectral index is independent of $w_\phi$ 
 for  $w_\phi \geq 1/9$ and in the 
 entire range, the spectral index varies from $-9/2$ to $-6$ for $0\leq w_{\phi}\leq 1$. So, unlike the IR divergence in large-scale 
 spectrum (\ref{longspectra}), there is no UV divergence in the 
 number density spectrum.
The spectra given in Eqs.~(\ref{longspectra}), (\ref{beta0.3f}), and (\ref{smallspectra}) are three important analytic results of this paper. In Figure \ref{fullspectrumfig}, we find a nice agreement between the numerically obtained small-scale spectra and the approximated analytical spectra given by Eqs. (\ref{beta0.3f}) and (\ref{smallspectra}) for different EoS. As already stated earlier, the small-scale modes never experience non-adiabatic evolution, resulting in a low spectral amplitude, $|\beta_k|^2\ll1$ for $k\gg\ke$. Therefore, to numerically study the spectral behavior in this regime, we rely upon computing the integral (\ref{Eq:mybeta1}) along with the background dynamical equations (\ref{finflaton}) and (\ref{Hubble}), which are utilized in studying the behavior of the time-dependent frequency term (\ref{Eq:mydot}). In the numerical analysis, we set the potential model parameter $\alpha=1$. However, the spectral behavior, anyway, is independent of the choice of this parameter.  Because of background expansion, the gravitational source term $\frac{a^{\prime\prime}}{a}$ in Eq.(\ref{Eqgravsource}) rapidly loses its efficiency for further production, thereby causing the stabilization of the amplitude of the integral (\ref{Eq:mybeta1}) within a few post-inflationary e-folding number. Now, it becomes interesting to compare our result with another approach, the Boltzmann one.
\section{Number density spectrum: Boltzmann versus Bogoliubov}\label{sec5}

 We start this discussion by exploring the perturbative production of massive scalar from the oscillating inflaton condensate which is treated as a classical field. We assume this condensate is homogeneous, decays perturbatively, and it follows the phase space distribution
\begin{align}\label{phasespaceinf}
     &f_{\phi}(k^{\prime},t)=(2\pi)^3n_{\phi}(t)\delta^{(3)}(\vec{k}^{\prime})\,,
 \end{align}
 where $n_{\phi}(t)$ is the instantaneous inflaton number density. In the perturbative approach, the 
 phase space distribution of the produced massive scalar field, $f_{\chi}$, is obtained by solving the Boltzmann transport equation
\begin{equation}\label{Boltzmaneq1}
\frac{\partial f_{\chi}}{\partial t}-H|\vec{p}|\frac{\partial f_{\chi}}{\partial |\vec{p}|}=C\big[f_{\chi}\big(|\vec{p}|,t\big)\big] \,,
 \end{equation}
where $\mathcal{C}\big[f_{\chi}\big]$ is the collision term given by
\begin{widetext}
\beq
\label{collision}
   C\big[f_{\chi}(|\vec{p}|,t)\big]=\frac{1}{2p^0}\sum_{\nu=1}^{\infty}\int\frac{d^3\vec{k}_{\nu}^{\prime}}{(2\pi)^3 n_{\phi}}\frac{d^3\vec{p^{\prime}}}{(2\pi)^3 p^{0\prime}}
   (2\pi)^4\delta^{(4)}\big(k_{\nu}^{\prime}-p-p^{\prime}\big) |\overline{\mathcal{M}_{\nu}}|^2_{\phi\rightarrow\chi\chi}\bigg[f_{\phi}(k^{\prime})\left(1+f_{\chi}(p)\right)\left(1+f_{\chi}(p^{\prime})\right)\bigg]\,,
\eeq
\end{widetext}
with $k_{\nu}^{\prime}=(E_{\nu},0)$, the four-momentum of the inflaton condensate and $E_{\nu}=\nu \omega$ denotes the energy of the inflaton condensate corresponding to the $\nu$-th mode of 
oscillation and $\omega$ is defined in Eq.~(\ref{fundafre}). 
Vanishing three momenta implies that inflaton decays to massive 
particles at its rest frame and $\overline{\mathcal{M}_{\nu}}$ 
is the transition amplitude in one oscillation for each 
oscillating field mode of $\phi$ from coherent condensate state $\ket{\phi}$ to two final particles state $\ket{\chi\chi}$. In the 
above expression of the collision term, the inverse decay term has been neglected satisfying the energy-momentum conservation. \textcolor{blue}{}\\ 
From energy-momentum conservation, we have
 \begin{align}\label{enermomencons}
&p=p^{\prime}=\frac{\nu \omega}{2}\sqrt{1-\frac{4m_{\chi}^2}{(\nu \omega)^2}}
 \end{align}
and considering very low-mass particles, $m_{\chi} \ll m_{\phi}$ or massless case, $m_{\chi}\approx 0$, we can write $p= p^{\prime}\approx \nu\omega/2$.\\

Now implementing (\ref{phasespaceinf}) and (\ref{collision}) in (\ref{Boltzmaneq1}), we get
\begin{align}\label{Boltzmaneq2}
    &\frac{\partial f_{\chi}}{\partial t}-H|\vec{p}|\frac{\partial f_{\chi}}{\partial |\vec{p}|}
    = \sum_{\nu=1}^{\infty}\frac{\pi|\overline{\mathcal{M}_{\nu}}|^2 }{2(p^0)^2}\delta\left(\frac{\nu \omega}{2}-p^0\right)\left(1+2f_{\chi}(p)\right)\nonumber\\
&\Rightarrow \frac{d}{d t}f_{\chi}(|\vec{p}|,t)=\sum_{\nu=1}^{\infty}\frac{2 \pi|\overline{\mathcal{M}_{\nu}}|^2 }{\nu^2 \omega^2}\delta\left(|\vec{p}|-\frac{\nu \omega}{2}\right)\nonumber\\
&\Rightarrow f_{\chi}(|\vec{p}|,t)= \sum_{\nu=1}^{\infty}\frac{2\pi}{\nu ^2}\int_{t_{\rm e}}^{t}\frac{|\overline{\mathcal{M}_{\nu}(t^{\prime})}|^2 }{\omega^2(t^{\prime})}\delta\left(|\vec{p}(t^{\prime})|-\frac{\nu \omega(t^{\prime})}{2}\right)dt^{\prime}\nonumber\\
&\Rightarrow f_{\chi}(|\vec{p}|,t)= \sum_{\nu=1}^{\infty}\frac{2\pi}{\nu ^2}\int_{t_{\rm e}}^{t}\frac{|\overline{\mathcal{M}_{\nu}(t^{\prime})}|^2 }{\omega^2(t^{\prime})}\delta\left(\frac{|\vec{p}(t)|a(t)}{a(t^{\prime})}-\frac{\nu \omega(t^{\prime})}{2}\right)dt^{\prime}
\,.
\end{align}
 In the above expression, for massless particles we have $p^0=|\vec{p}|=\big(\nu \omega(t)\big)/2$. As here we are talking 
 about gravitational spectra, the condition, $f_{\chi}(p)\ll 1$, is essentially valid. In the second line of 
 Eq.~(\ref{Boltzmaneq2})  we have assumed $\big(1+2f_{\chi}(p)\big)\approx 1$ and in the final line, the product $|\vec{p}(t)|a(t)$ is the comoving momentum. 
 
To evaluate the delta function in the integral (\ref{Boltzmaneq2}), we introduce a time $\hat{t}$ between the two limits $t_{\rm e}$ and $t$ which satisfies the relation
\begin{equation}\label{tcap}
   \frac{a(t)}{a(\hat{t})}=\frac{\nu \omega(\hat{t})}{2|\vec{p}(t)|} \,.
\end{equation}

 Here we are considering gravitons ($h_{\mu\nu}$) mediated production of a minimally coupled scalar field ($\chi)$ from the inflaton background (see \cite{ Clery:2021bwz} for a detailed analysis), and do not consider any other interaction between inflaton and scalar field. For the minimal case, the gravitational interaction gives the following form of the interaction Lagrangian.
\begin{equation}\label{minint}
 \mathcal{L}_{\rm int}^{}= -\frac{ h^{\mu\nu}}{M_P}\left(T^{\phi}_{\mu\nu}+T^{\chi}_{\mu\nu}\right)   
\end{equation}
where $T^{\phi}_{\mu\nu}$, and $T^{\chi}_{\mu\nu}$ are the energy-momentum tensors of the inflaton and scalar field, respectively.
Based on the above interaction Lagrangian, we calculate the expression of the transition amplitude $\mathcal{M}_{\nu}$ as follows:  \cite{Clery:2021bwz,Garcia:2022vwm, Chakraborty:2025oyj} 
\begin{align}\label{transitionamp1}
    \mathcal{M}_{\nu}&=-\frac{1}{M_{P}^2}V(\Phi)\mathcal{P}^{2n}_{\nu}\left(1+\frac{2m_{\chi}^2}{(\nu \omega)^2}\right)\nonumber\\
    &=-\frac{1}{M_{P}^2}\rho_{\phi}\mathcal{P}^{2n}_{\nu}\left(1+\frac{2m_{\chi}^2}{(\nu \omega)^2}\right) \, .
\end{align}
Now, if we take into account the symmetry factor associated with two identical final states, then the average transition probability amplitude for massless(or, very low mass) final state particles becomes
\begin{equation}\label{transitionamp2}
\sum_{\nu=1}^{\infty} |\overline{\mathcal{M}_{\nu}}|^2 =\frac{1}{2}\times \sum_{\nu=1}^{\infty}\frac{\rho^2_{\phi}}{M_{P}^4}|\mathcal{P}^{2n}_{\nu}|^2\,,
\end{equation}
Using Eqs.(\ref{tcap}) and (\ref{transitionamp2}) in (\ref{Boltzmaneq2}) we can then evaluate the phase space distribution function for general EoS in the low mass limit ($m_{\chi}\approx 0 $)
\begin{align}
   f_{\chi}(p,t)
   &= \sum_{\nu=1}^{\infty}\frac{2\pi|\mathcal{P}^{2n}_{\nu}|^2}{\nu^3}\frac{\rho^2_{\phi}(\hat{t})}{M^4_{P}\omega^2(\hat{t})}\frac{\theta(t-\hat{t})\theta(\hat{t}-t_{\rm e})}{\left|-\omega(\hat{t})H(\hat{t})-\dot{\omega}(\hat{t})\right|}\nonumber\\
   &=\sum_{\nu=1}^{\infty}\frac{2\pi|\mathcal{P}^{2n}_{\nu}|^2}{\nu^3}\frac{\rho^2_{\phi}(\hat{t})}{M^4_{P}\omega^3(\hat{t})H(\hat{t})}\frac{\theta(t-\hat{t})\theta(\hat{t}-t_{\rm e})}{|3w_{\phi}-1|}\nonumber\\
&=\sum_{\nu=1}^{\infty}\frac{2\pi|\mathcal{P}^{2n}_{\nu}|^2}{\nu^3 |3w_{\phi}-1|}\frac{\rho^2_{\phi}(t)}{M^4_{P}\omega^3(t)H(t)}\left(\frac{2p(t)}{\nu \omega(t)}\right)^{\frac{9(w_{\phi}-1)}{2(1-3w_{\phi})}}\nonumber\\
&\times\theta(t-\hat{t})\theta(\hat{t}-t_{\rm e})\nonumber\\
  &=\frac{\pi}{4M_{P}^4|3w_{\phi}-1|}\frac{\rho^2_{\text{e}}}{H_{\text{e}}(m_{\phi}^{\text{e}}\bar{\alpha})^3}\left(\frac{m^{\rm e}_{\phi}\bar{\alpha}}{p(t)(a/a_{\rm e})}\right)^{\frac{9(1-w_{\phi})}{2(1-3w_{\phi})}}\nonumber\\
   \begin{split}
&\times\sum_{\nu=1}^{\infty}|\mathcal{P}^{2n}_{\nu}|^2\left(\frac{\nu}{2}\right)^{\frac{3(1+3w_{\phi}}{2(1-3w_{\phi})}}\theta(t-\hat{t}) \theta(\hat{t}-t_{\rm e})
   \end{split}\nonumber\,,\\
     f_{\chi}(k,a)&=\frac{9\pi}{4}(\bar{\alpha})^{\frac{3(1+3w_{\phi})}{2(1-3w_{\phi})}}\left(\frac{H_{\text{e}}}{m_{\phi}^{\text{e}}}\right)^3\left(\frac{a_{\rm e} m_{\phi}^{\text{e}}}{k}\right)^{\frac{9(1-w_{\phi})}{2(1-3w_{\phi})}}\nonumber\frac{1}{|(3w_{\phi}-1)|}\nonumber \\
    \begin{split}
    &  \times\sum_{\nu=1}^{\infty}|\mathcal{P}^{2n}_{\nu}|^2\left(\frac{\nu}{2}\right)^{\frac{3(1+3w_{\phi})}{2(1-3w_{\phi})}}
\theta\left(\left(\frac{2k}{\nu a_{\rm e} m_{\phi}^{\text{e}}\bar{\alpha}}\right)^{\frac{1}{1-3w_{\phi}}}-1 \right)
\end{split}\nonumber\\
\begin{split}
&
\theta\left(\left(\frac{a}{a_{\rm e}}\right)\left(\frac{2k}{\nu a_{\rm e} m_{\phi}^{\text{e}}\bar{\alpha}}\right)^{\frac{1}{3w_{\phi}-1}}-1 \right)\,.
    \end{split}
   \end{align}
and $\bar \alpha$ is defined in Eq.~(\ref{Eq:alphabar}).
Using the relations 
\begin{align}
& \rho_{\phi}(\hat{t})=\rho_{\phi}(t)\left(\frac{a(\hat{t})}{a(t)}\right)^{-3(1+w_{\phi})};H(\hat{t})=H(t)\left(\frac{a(\hat{t})}{a(t)}\right)^{-\frac{3 (1+w_\phi)}{2}}~~\nonumber\\
& \omega(\hat{t})=\omega(t)\left(\frac{a(\hat{t})}{a(t)}\right)^{-3w_{\phi}}~~;~~ \rho^2_{\text{e}}=9M_{P}^4H_{\text{e}}^4
\end{align}
together with $(a(\hat{t})/a(t))=\Big(2p(t)/\nu \omega(t)\Big)^{\frac{1}{1-3w_{\phi}}}$ which is deduced from  Eq.~(\ref{tcap}), we finally obtain the distribution function as follows
\begin{widetext}
\begin{align}\label{Boltzmaneqf}
 f^{\wre\neq 1/3}_{\chi}(k,a) &= \frac{9\pi}{4|3w_{\phi}-1|}\left(\frac{\bar{\alpha}m_{\phi}^{\text{e}}}{H_{\text{e}}}\right)^{\frac{3(1+3w_{\phi})}{2(1-3w_{\phi})}}\left(\frac{\ke}{k}\right)^{\frac{9(1-w_{\phi})}{2(1-3w_{\phi})}}\sum_{\nu=1}^{\infty}|\mathcal{P}^{2n}_{\nu}|^2\left(\frac{\nu}{2}\right)^{\frac{3(1+3w_{\phi})}{2(1-3w_{\phi})}}
\theta\left(\left(\frac{2k}{\nu a_{\rm e} m_{\phi}^{\text{e}}\bar{\alpha}}\right)^{\frac{1}{1-3w_{\phi}}}-1 \right)\nonumber\\
&\times\theta\left(\left(\frac{a}{a_{\rm e}}\right)\left(\frac{2k}{\nu a_{\rm e} m_{\phi}^{\text{e}}\bar{\alpha}}\right)^{\frac{1}{3w_{\phi}-1}}-1 \right)\,
\end{align}
\end{widetext}

In the above equation (\ref{Boltzmaneqf}), $(p(t)a(t)=k)$ is the comoving momentum of the produced particles. 
 To the leading order, this is our main result obtained from the Boltzmann equation, which describes the spectral behavior of 
 the produced particles for the sub-horizon modes if $w_\phi \neq 1/3$. From the 
 distribution function (\ref{Boltzmaneqf}), we indeed recover the well known spectral nature $f_{\chi}\propto k^{-9/2}$ for $w_{\phi}=0$. If we further compare this with the resulting 
 distribution function Eqs.~(\ref{beta0.3f}) and (\ref{smallspectra}) obtained by the Bogoliubov approach, one 
 clearly sees that both matches for a matter like reheating equation of state and also the higher equation of state
 (see Fig.(\ref{fullspectrumfig})).
It has already been shown in the Bogoliubov method that the 
spectral index for UV modes varies from $-9/2$ to $-6$ for $0 \leq w_{\phi}\leq 1$, therefore, in the computation of energy 
density, no UV divergence occurs. But, the spectrum (\ref{Boltzmaneqf}) having spectral index $\big(9(w_{\phi}-1)/2(1-3w_{\phi})\big)$ appears to be UV 
divergent both in energy density and particle number density for EoS $w_{\phi}>1/3$  which is seemingly in sharp contradiction with 
the spectrum obtained in the non-perturbative method earlier. 
Surprisingly, the computation of the full sum containing the Fourier components and two Heaviside functions in the Eq.~(\ref{Boltzmaneqf}) finally generates a convergent spectral nature which agrees well with the Bogoliubov spectrum in the UV regime, as we can see in Fig.~(\ref{fullspectrumfig}) for $\wre=0, ~1/2,~2/3$. In this regard, a few important points are mentioned by analyzing the massless distribution function in Eq.~(\ref{Boltzmaneqf}). 
\begin{itemize}
    \item \underline{For $0<w_{\phi}<1/3$:} In this EoS range, $(1-3w_{\phi})>0$ and this makes the index in the first Heaviside function (theta function) of (\ref{Boltzmaneqf}) always positive. This makes us inclined towards a high-frequency regime, $k/a_{\rm e}m^{\rm e}_{\phi}>1$, satisfying the condition $(k/a_{\rm e}m^{\rm e}_{\phi})\geq \nu\bar{\alpha}/2$ to have a non-vanishing value from the first Heaviside function.\\
    On the other hand, in the second Heaviside function, the index alters its sign and gets negative in this EoS range. Therefore, in the high-frequency regime, we need to go to a larger time scale, $a(t)/a_{\rm e} \gg 1$, to satisfy the second Heaviside function. From (\ref{tcap}), such instants say $\hat{t}$ are calculated as $(a(\hat{t})/a_{\rm e})=\big(2k/\nu a_{\rm e}m^{\rm e}_{\phi}\bar{\alpha}\big)^{1/(1-3w_{\phi})}$ and these instants are nothing but the stationary points as interpreted in the Bogoliubov approach and they indeed satisfy this second Heaviside function in high-frequency regime. For the stationary points, we get another condition $(k/a_{\rm e}m^{\rm e}_{\phi})\leq (\nu\bar{\alpha}/2)(a(t)/a_{\rm e})^{(1-3w_{\phi})}$.
    \item \underline{For $w_{\phi}>1/3$:} In this EoS range, spectrum index $\big(9(w_{\phi}-1)/2(1-3w_{\phi})\big)$ starts diverging in UV regime. However, to satisfy two Heaviside functions simultaneously, we get more constrained. Now the index of the first theta function gets negative, which eventually gives the condition $\nu\bar{\alpha}/2\geq (k/a_{\rm e}m^{\rm e}_{\phi})$ to satisfy the first theta function. To satisfy the second Heaviside function, we need to choose the time instants satisfying the condition, $\left(a(t)/a_{\rm e}\right)\geq \left(\nu a_{\rm e}m^{\rm e}_{\phi}\bar{\alpha}/2k\right)^{\frac{1}{(3\wre-1)}}$.
    The condition $\nu\bar{\alpha}/2\geq (k/a_{\rm e}m^{\rm e}_{\phi})$ actually discretizes the computation of the sum in Eq.(\ref{Boltzmaneqf}) and, for an arbitrary high-frequency mode, we cannot take the sum from any order of Fourier components. If we go on increasing the ratio $\left(k/a_{\rm e}m^{\rm e}_{\phi}\right)$, we must take higher and higher order of Fourier components $\mathcal{P}^{2n}_{\nu}$. Again with the increase of order $\nu$, $\mathcal{P}^{2n}_{\nu}$ drops rapidly, and after first few terms of the series, it becomes vanishingly small which in turn restricts the maximum range of high-frequency mode up to which one can reach. Beyond that momentum range, the Fourier sum in (\ref{Boltzmaneqf}) will have a vanishing contribution. Although there is a diverging spectral index in this EoS range, $k^{\frac{9(w_{\phi}-1)}{2(1-3w_{\phi})}}$, the associated Heaviside functions interestingly maintain the convergence of the spectrum in the high-frequency limit, and the UV spectrum follows $k^{-6}$ behavior regardless of the EoS as predicted by non-perturbative analysis in the previous section. 
\end{itemize}


 For the specific case of $w_\phi = 1/3$, we have that 
\begin{equation}
f_{\chi}(|\vec{p}|,t)= \sum_{\nu=1}^{\infty}\frac{2\pi}{\nu ^2}\int_{t_{\rm e}}^{t}\frac{|\overline{\mathcal{M}_{\nu}(t^{\prime})}|^2 }{\omega^2(t^{\prime})}\frac{a(t')}{a(t)}\delta\Big(|\vec{p}(t)|-\frac{\nu \omega(t)}{2}\Big)dt^{\prime}
\end{equation}
where now the Dirac delta is independent of the variable of integration $t'$ due to the same redshift of momenta and inflaton frequency, $\omega(t')=\omega(t)\left(a(t)/a(t')\right)$. Thus, it can be brought out of the integral and we are left with the integral from $t_{\rm e}\rightarrow t$ of the quantity
\begin{equation}
    \int_{t_{\rm e}}^t\frac{a(t')}{a(t)}\frac{|\overline{\mathcal{M}_{\nu}(t^{\prime})}|^2 }{\omega^2(t^{\prime})} dt' =|\mathcal{P}^{2n}_\nu|^2\int_{t_{\rm e}}^t \frac{a(t')}{a(t)}\frac{\rho_\phi^2(t')}{2M_P^4 \omega^2(t')} dt' \, .
\end{equation}
Considering the background evolution of the inflaton energy density for $w_\phi = 1/3$ we obtain then the following Boltzmann distribution function
\begin{widetext}
\begin{align}\label{boltzspec0p33}
\boxed{
   f_{\chi}^{w_\phi=1/3}(k,a) = 3\pi\left(\frac{H_e}{m_\phi^e \bar{\alpha}}\right)^2\left[1-\left(\frac{a_e}{a}\right)^3\right]
   \sum_{\nu = 1 }^{+\infty} \frac{|\mathcal{P}^{2n}_\nu|^2}{\nu^2}\delta\left(\frac{k}{\ke}-\frac{\nu \bar{\alpha}m_\phi^e}{2H_e}\right) \, .
   }
\end{align}
\end{widetext}
It corresponds to a series of peaks located at different momenta $(1/2)\nu m_\phi^e\bar{\alpha}$ as previously described in \cite{Garcia:2023eol}
for the distribution function of inflaton quanta produced by inflaton self-interactions in a quartic potential, for which $w_\phi=1/3$. The appearance of these discrete delta peaks in the momentum space causes the initial rapid fluctuations in the Boltzmann spectrum for $\wre=1/3$ in Fig.(\ref{fullspectrumfig}). One should note, however, that in the context of self-interactions, the occupation number was found to rapidly grow after inflation, scaling as $\propto a(t)/a_e$ for each peak, signaling the breakdown of the perturbative Boltzmann picture rapidly after inflation. In the case of the gravitational production of quanta, we obtain here that the spectrum is fully determined by what happens at the end of inflation, and the occupation number for each mode is not growing as a function of time, allowing us to consider the Boltzmann computation even at late times for $w_\phi=1/3$. \\
To compare the nature of the phase-space distribution function or number density spectrum obtained in two different approaches, we present their behavior in Figure (\ref{fullspectrumfig}) for four different EoS $w_{\phi}=0,~1/3,~ 1/2,~ 2/3$, where we considered a maximum Fourier scale $\nu_{\rm max}= 2000$ for the Boltzmann numerical computation. We observe that the Bogoliubov approach and the Boltzmann approach agree quite well on the shape of the spectrum, as well as with our analytical approximation for all four EoS, and this accordance is true for any higher EoS $\wre\geq 1/3$.


\section{Reheating dynamics through gravitational production }\label{sec6}

 In this section, we shall investigate the reheating process through pure gravitational production considering the full 
 number density spectrum $|\beta_k|^2$ associated with long and short-wavelength modes. In this estimation, we make use of the analytic expressions of the spectrum computed in 
  section \ref{sec4}. To determine the reheating temperature through the gravitational excitation of scalar modes, we assume the thermalization of the produced modes once they become subhorizon (particle states). This approximation of fast equilibration may not be valid depending on the additional interactions to consider among the relativistic particles produced after inflation. Several studies have been focusing on the problem of reaching chemical and kinetic equilibrium for a simple relativistic sector produced during preheating via classical simulations \cite{Felder:2000hr, Micha:2002ey, Micha:2004bv, Yoon:2024pbz} or through perturbative quantum processes during reheating \cite{Davidson:2000er, Kurkela:2011ti, Harigaya:2013vwa, Mukaida:2024jiz}. Here, we consider the possibility of producing the light Standard Model (SM) quanta from the inflaton through minimal gravitational interaction, especially in the form of relativistic Higgs bosons. In this case, the large number of degrees of freedom in the SM and the sizable gauge interactions within this sector are expected to efficiently bring
the produced relativistic particles into a hot thermal bath. We do not investigate further the complexity of non-instantaneous thermalization during and after reheating, which requires dedicated numerical analysis. Still, we assume we can associate a temperature with the produced relativistic particles at the end of the reheating era.
 
Our present analysis clearly shows the requirement of a  high post-inflationary EoS
(or, some stiff 
matter-dominated phase) to achieve successful reheating through pure gravitational production. It also shows the sensitive dependence of reheating temperature on 
  the inflationary energy scale through the variation of the 
  parameter $\alpha$ in the potential (\ref{alphaattractor}). 
  
  We have obtained the power-law-type behavior of 
  particle number density spectrum in both IR and UV regimes for 
  different EoS and the detailed investigation reveals that minimum 
  reheating temperature that is BBN (\textit{Big Bang nucleosynthesis}) bound,  $\Trh \sim 1~\rm MeV $, is achieved around $w_{\phi}=0.6$ for $\alpha=1$. If we keep on increasing 
  the parameter value $\alpha$, we keep on getting lower and 
  lower reheating temperatures at higher EoS. The more the 
  value of EoS $w_{\phi}$ that is closer to the kination 
  regime $(w_\phi\rightarrow1)$, the higher the reheating temperature. Indeed, in the gravitational reheating scenario, in the absence of any such specific decay 
  channel for the inlfaton, background inflaton energy density scales as $\rho_{\phi}\propto a^{-3(1+w_{\phi})}$ whereas produced 
  massless fluctuation being radiation scales as $\rho_{R}\propto a^{-4}$. In order to successfully reheat the 
  universe, $\rho_{\phi}$ and $\rho_R$ must reach their equality and this essentially requires EoS $w_{\phi}> 1/3$ in 
  the pure gravitational scenario. Therefore, simple energy scaling confirms the requirement of higher EoS for gravitational reheating to happen. Now, 
  combining UV and IR modes and using the spectrum given in Eq.(\ref{longspectra}) and (\ref{smallspectraappen}), the total comoving energy density of produced particles can be computed as follows
\begin{widetext}
\begin{align}\label{comenergyuvir1}
    \rho_{R}a^4&=\int_{k_{\rm RH}}^{\ke} \frac{k^3}{2\pi^2} |\beta_k|_{\rm IR}^2 dk+ \int_{\ke}^{k_{\rm Planck}} \frac{k^3}{2\pi^2} |\beta_k|_{\rm UV}^2 dk \nonumber\\
\Rightarrow \rho_{R}a^4&=\frac{\mathcal{D}\ke^4 }{4\pi^3(1-2\bar{\nu})}\bigg(1-\left(\frac{ k_{\rm RH}}{\ke}\right)^{(1-2\bar{\nu})}\bigg)+\frac{a_{\rm e}^6\sum\Big[ {\cal N}_0 + {\cal N}_1 +{\cal N}_2+{\cal N}_3+{\cal N}_8+{\cal N}_9\Big]^2}{64\pi^2 (f(w_{\phi}))^2\ke^2}\times \bigg(1-\left(\frac{\ke}{k_{\rm Planck}}\right)^2\bigg)\nonumber\\
\Rightarrow \rho^{\rm com}_R(\alpha,~\wre)&\approx \frac{\mathcal{D}(\wre)\He^4(\alpha,~\wre)} {4\pi^3(1-2\bar{\nu}(\wre))}+\frac{\sum\Big[ {\cal N}_0 + {\cal N}_1 +{\cal N}_2+{\cal N}_3+{\cal N}_8+{\cal N}_9\Big]^2}{64\pi^2 (f(w_{\phi}))^2\He^2(\alpha,~\wre)}\nonumber\\
\Rightarrow \rho^{\rm com}_R(\alpha,~\wre)&\approx \frac{\mathcal{D}(\wre)(3\wre+1)\He^4(\alpha,~\wre)} {8\pi^3(3\wre-1)}+\frac{9(1+\wre)^2\sum\Big[ {\cal N}_0 + {\cal N}_1 +{\cal N}_2+{\cal N}_3+{\cal N}_8+{\cal N}_9\Big]^2}{64\pi^2 (1+3\wre)^2\He^2(\alpha,~\wre)}
\end{align}   
 \end{widetext}
In the expression above, the first part is associated with the large-scale (IR) 
contribution and the second part is associated with the small-scale (UV) contribution, where the coefficients $\mathcal{N}_i$ are functions of the model parameter $\alpha$ and the EoS $\wre$ as defined earlier and in the Appendix \ref{appenA}. Another EoS dependent function $\mathcal{D}$ is defined in (\ref{D}). Here $k_{\rm RH}=a_{\text{RH}} H_{\text{RH}}$ is the scale that entered the horizon at the end of reheating, and $a_{\text{RH}}, H_{\text{RH}}$ are 
the scale factor and Hubble scale at the end of reheating. Furthermore, in the comoving energy density expression, we take the 
IR cut-off at the scale $k_{\rm RH}$ and we take the UV cut-off at the Planck energy scale, $k_{\rm Planck}$. It is important to note that while defining any physical quantity like energy density or temperature associated with fluctuations, we shall consider solely the contribution of the causal modes, which are well inside the horizon at the time of computing the quantity of interest. The highest accessible scale in the finite time scale of reheating is $\kre$, and the modes in the range $\kre\leq k\leq\ke$ are well inside the horizon at the end of reheating, hence contributing to the total energy density of produced particles \cite{Chakraborty:2025oyj}. Therefore, $\kre$ is considered to be the IR limit in the computation of $\rho^{\rm com}_R$ at the time of reheating in Eq.(\ref{comenergyuvir1}).
The very nature of the UV spectrum shows no small-scale divergence at all in the computation of the energy density for the 
entire range of EoS $0\leq w_{\phi}\leq 1$, whereas the very 
nature of the IR spectrum (see (\ref{longspectra})) suggests 
a large-scale divergence in the total energy density for $w_{\phi}< 1/3$, and a 
logarithmic divergence for $ w_{\phi}=1/3$. However, in the entire range $w_{\phi}>1/3$, there is no IR divergence in the energy 
density spectrum. 
Therefore, in the gravitational reheating scenario with $\wre\geq 3/5$, we are free from both IR and UV divergence of the comoving energy spectra. As $k_{\rm Planck} \gg \ke$ and $k_{\rm RH} \ll \ke$, in both the cases for $w_{\phi}>1/3$, 
the significant 
contribution in energy comes around the scale, which leaves the Hubble horizon at the inflation end, which is $k=\ke$.\\ 
As we are dealing with pure gravitational production here,  the evolution of background inflaton energy density will mostly be affected by background expansion rather than the backreaction of produced fluctuations, as mentioned before. Therefore, background energy density evolves as 
\begin{equation}\label{energyinflaton}
   \rho_{\phi}(a)\simeq(3M_{ P}^2H_{\text{e}}^2)\left(\frac{a}{a_{\text{e}}}\right)^{-3(1+w_{\phi})} 
\end{equation}
Different spectral behavior causes different reheating e-folding number $N_{\rm RH}\equiv \ln{\big(\frac{a_{\rm RH}}{a_{\rm e}}\big)}$ for different EoS. In order to obtain $N_{\rm RH}$ for different $w_{\phi}$, the condition $\rho_{\phi}(a)=\rho_R(a)$ has to be met at some point, say, at $a_{\rm RH}$ during the reheating phase. We have,
\begin{equation}\label{are}
    \left(\frac{a_{\rm RH}}{a_{\rm e}}\right)=\left(\frac{3M_{ P}^2H_{\text{e}}^2}{\rho_R^{\rm com}}\right)^{\frac{1}{(3w_{\phi}-1)}}
\end{equation}

\begin{figure*}
\begin{center}
\includegraphics[width=0.45\linewidth]{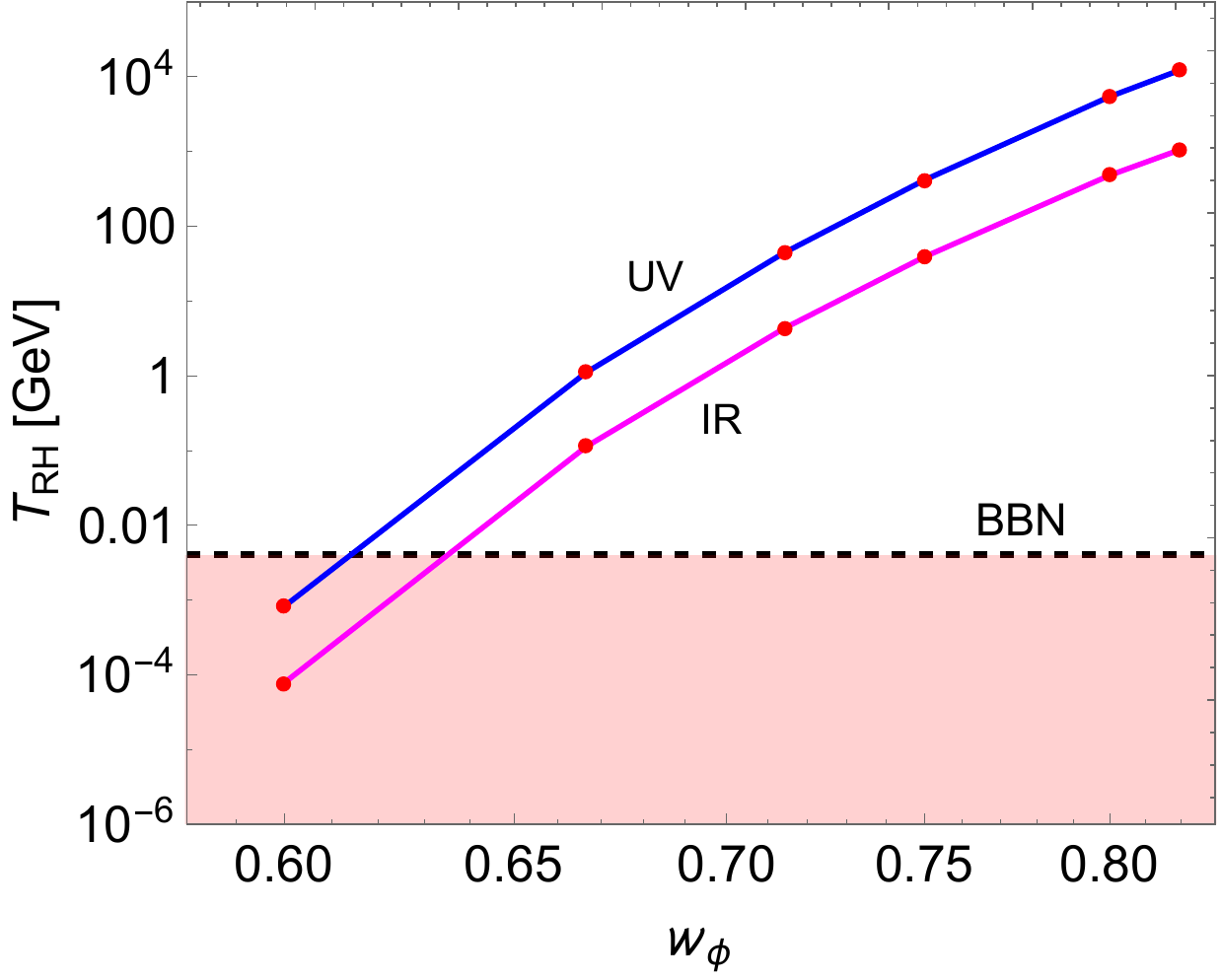}
\caption{\textit{
The figure represents the individual contribution of IR and UV components of $\rho_{\rm R}^{\rm com}$ in Eq.(\ref{comenergyuvir1}) to the reheating temperature for varying EoS. The BBN temperature here is taken to be} 4 \textit{MeV}.
}
\label{reheattempUVIRfig}
\end{center}
\end{figure*}

Finally, combining both IR and UV scale contributions, we end up having the expression of reheating temperature achieved in the pure gravitational reheating scenario as  
\begin{align}\label{reheattemp}
   T_{\text{RH}}
   &=\left(\frac{30}{g_{\text{RH}}\pi^2}\right)^{\frac{1}{4}}\big(3M_{ P}^2H_{\text{e}}^2\big)^{\frac{1}{1-3w_{\phi}}}\big(\rho^{\rm com}_R\big)^{-\frac{3(1+w_{\phi})}{4(1-3w_{\phi})}}
\end{align}
where $g_{\rm RH}$ is the total relativistic degrees of freedom at the time of reheating, which is given by $g_{\rm RH}=106.75$ in the SM for $\Trh \gtrsim 1~\rm GeV$.
Eq.~(\ref{reheattemp}) along with (\ref{comenergyuvir1}) are two important results of this paper giving us a particular reheating temperature corresponding to a specific reheating EoS {\it within the Bogoliubov approach}. Our detailed numerical results of reheating dynamics based on this non-perturbative formalism compared with the Boltzmann approach are given in Table \ref{tab1} for different values of $w_\phi$.
We note that both approaches agree quite well numerically. The slight discrepancy in the temperature prediction is caused by the addition of the IR contribution to the total energy density as well as the slightly higher spectral amplitude obtained in the non-perturbative Bogoliubov treatment, compared to the Boltzmann one, around the scale $k\sim\ke$, for any $\wre>1/3$.
Now, to understand the individual contributions of the IR and UV 
components of radiation energy density
$\rho_{\rm R}^{\rm com}$ from Eq.(\ref{comenergyuvir1}), we 
also calculated the reheating temperatures for these two components separately for varying EoS as shown in Fig.(\ref{reheattempUVIRfig}). Evidently, in this non-perturbative minimal reheating scenario, the small-scale(UV) contribution in the energy density is slightly larger than the large-scale(IR) contribution causing the difference in the reheating temperatures as computed through UV and IR modes separately in Fig.(\ref{reheattempUVIRfig}). 
\begin{table}[h!]
 \caption{\textit{Variation of reheating temperature with various EoS for $\alpha=1$. Here we have given both Boltzmann and Bogoliubov predictions for comparison.}}
 \vspace{2ex}
 \centering
\begin{tabular}{|c|c|c|c|}
\hline
\multicolumn{2}{|c|}{Bogoliubov} & \multicolumn{2}{|c|}{Boltzmann}  \\
\hline
\cline{1-4}
 $w_{\phi}$ & $\text{T}_{\text{RH}}~(\text{GeV})$& $w_{\phi}$ &  $\text{T}_{\text{RH}}~(\text{GeV})$\\
\hline
 3/5 & $1.12\times 10^{-3}$ & 3/5 &$9.40\times10^{-4}$ \\
 \hline
 2/3 & 2.17  & 2/3 &$0.97$ \\
 \hline
 5/7 & 58.26 & 5/7 & $11.76$ \\  
 \hline
 3/4 & $3.54\times 10^{2} $ &   3/4 & $ 2.42\times 10^{2}$ \\
 \hline
 4/5 & $5.84\times 10^{3}$ & 4/5 &$2.54\times10^{3}$ \\
 \hline
\end{tabular}
\label{tab1}
\end{table}

\section{ Conclusion}\label{sec7}

The reheating epoch is the least explored epoch in the entire cosmic history to date because of the absence of any direct observational evidence. This phase connects the early inflation and the thermal radiation-dominated phase of the universe by setting the necessary initial conditions for hot \textit{Big Bang Nucleosynthesis} (BBN). All the elementary particles populating the present universe were created in this epoch. Therefore, the exploration of this phase is unavoidable indeed.

In this work, we have computed the long and short-wavelength spectra of a massless scalar fluctuation during inflation and post-inflationary reheating employing the non-perturbative Bogoliubov treatment in the pure gravitational reheating 
scenario. In the long-wavelength regime, we found that 
in the case of a transition from de Sitter inflation to general reheating EoS $\wre$, the spectral index varies from -6 to -3 for $0\leq\wre\leq 1$. The lower the reheating EoS $\wre$, the heavier the red tilt of the spectrum in the IR regime. This IR 
behavior of the spectra is obtained within a 
purely non-perturbative treatment as the perturbative (or Boltzmann) 
approach fails to track the super-horizon evolution of the 
fluctuation. 

We have also studied the mass-breaking effect or 
the departure of the large-scale spectra from the massless 
limit. We have analytically computed the relationship between 
field mass ($m_{\chi}$) and the IR scale ($k_m$) below which all the 
modes suffer from the finite mass effect for any general reheating EoS. For any $m_{\chi}/\He\gtrsim 3/2$, the large-scale spectrum is found to experience an exponential 
mass-suppression effect. Interestingly, we find that for the masses $m_{\chi}/\He \ll 3/2$, the $k^3\big{|}\beta_k\big{|}^2$ spectral shape always remains flat irrespective of the post-inflationary 
EoS. Our numerical result also supports this analytical finding.\\
However, for the short-wavelength modes, we find the 
post-inflationary inflaton oscillation plays a very important 
role in computing the sub-horizon spectrum in the non-
perturbative treatment. Our study reveals that the product of 
an oscillatory term and a non-oscillatory term in the 
expression of $|\beta_k|^2_{\rm UV}$ results in the appearance 
of an interference term which nicely explains the origin of the 
momentum-space oscillations in the high-frequency regime of the 
spectrum as we can see in Fig.~(\ref{fullspectrumfig}). For $\wre=0$, we recover the known UV spectral index to be -9/2 with small oscillations over it. Interestingly, we find the EoS 
$\wre=1/9$, at which the dominant spectral index turns out to be -6, and for any $1/9\leq \wre\lesssim 1$, the spectral index remains -6 independent of the choice of $\wre$. Therefore, in the entire range 
$0\leq \wre\lesssim 1$, the UV spectral index varies from $-9/2$ to -6. Unlike the long-wavelength regime, there is no UV divergence in the short-wavelength regime of the number density spectrum. We have also found an agreement in the spectral behavior between two approaches, using the Bogoliubov formalism or solving the associated Boltzmann equation, in the UV regime for any EoS $0\leq\wre\lesssim 1$.\\\\
In the present study, we have also looked at how to achieve a successful reheating scenario from the non-perturbative 
approach. Such a gravitational reheating scenario is drawing much attention because of its simpler mechanism where a 
dynamical background itself suffices the conditions for successful reheating without the requirement of any complicated 
non-minimal coupling between produced fluctuation and inflaton or with gravity. In our work, we have studied this reheating scenario employing a non-perturbative Bogoliubov approach.
We presented our result in Section \ref{sec6}, summarizing the variation of the reheating temperature with $w_\phi$ in Table \ref{tab1}. Our analysis shows that maintaining a reheating temperature above the BBN energy scale requires \( w_{\phi} \gtrsim 0.6 \). Evidently, to achieve radiation domination through purely gravitational production, the Universe has to transit 
to a higher post-inflationary EoS. 
It is 
interesting to note that there is a one-to-one correspondence between 
reheating temperature and reheating EoS, so a particular EoS can 
uniquely specify a particular $T_{\text{RH}}$ as given in Table \ref{tab1}. However, such a gravitational reheating scenario generally faces the challenge of overproducing primary gravitational waves (PGWs) \cite{Barman:2022qgt, Barman:2024slw,Chakraborty:2023ocr, Barman:2023ktz}. As well, the generation of light scalar perturbations from inflation and reheating may suffer from the problem of a too large isocurvature power spectrum $\mathcal{P}_{S}(k)$ for these modes. These isocurvature perturbations are tightly constrained by \textit{Planck} data at CMB horizon crossing scale, $k_{\ast}/a_{0}=0.05 ~\text{Mpc}^{-1}$, with $a_0$ be the present-day scale factor. In this work, we are focusing on the gravitational production of a scalar fluctuation minimally coupled to gravity, $\xi=0$. In several recent studies \cite{Garcia:2023awt, Garcia:2023qab, Kolb:2023ydq, Chakraborty:2024rgl, Choi:2024bdn, Garcia:2025rut}, it has been clearly pointed out that for $\wre=0$, a massless scalar fluctuation must have a lower limit of non-minimal coupling strength, $\xi\gtrsim 0.027$ to prevent the overproduction of isocurvature fluctuation at the CMB scale respecting the current isocurvature bound $\mathcal{P}_{S}(k_{\ast})<8.3\times 10^{-11}$ \cite{Planck:2018jri, Planck:2018vyg}. This rules out the possibility of a minimally coupled scalar field for $\wre=0$ due to the heavily red-tilted large-scale spectrum, $|\beta_k|^2_{\rm IR}\propto k^{-6}$. On the contrary, for $\wre>1/3$, the appearance of strong post-inflationary \textit{tachyonic instability} beyond a certain coupling strength ($\xi>1/6$) for the large scales imposes upper bounds on $\xi$ respecting the isocurvature constraint at the CMB scale. These possibilities are explored in the recent studies \cite{Chakraborty:2024rgl, Chakraborty:2025oyj} quite extensively. It has also been found that, as one approaches $\wre=1/3$, the value of $\xi$ remains unconstrained subject to this large-scale isocurvature bound. We have not, anyway, explored these issues in the present work and leave it for future analysis.

\vspace{5px}
\section*{Acknowledgments}
The authors thank Kunio Kaneta, Mathieu Gross and Jong-Hyun Yoon for extremely valuable discussions and substantial numerical support.
AC would like to thank the Ministry of Human Resource Development, Government of India (GoI), for financial assistance. The work of SC is partially supported by the Collaborative Research Center SFB1258, by the Deutsche Forschungsgemeinschaft (DFG, German Research Foundation) under Germany's Excellence Strategy - EXC-2094 - 390783311 and also by a TUM University Global Postdoc Fellowship. MRH acknowledges ISI Kolkata for providing financial support through the Research Associateship. DM wishes to acknowledge support from the Science and Engineering Research Board~(SERB), Department of Science and Technology~(DST), Government of India~(GoI), through the Core Research Grant CRG/2020/003664.
This work was made possible by Institut Pascal at Université Paris-Saclay during the Paris-Saclay Astroparticle Symposium 2023 and 2024, with the support of the P2IO Laboratory of Excellence (program “Investissements d’avenir” ANR-11-IDEX-0003-01 Paris-Saclay and ANR-10-LABX-0038), the P2I axis of the Graduate School of Physics of Université Paris-Saclay, as well as IJCLab, CEA, IAS, OSUPS, APPEC, and the IN2P3 master projet UCMN.
\onecolumngrid
\appendix
\section{\textbf{Sub-horizon or UV($k>a_{\rm e}H_{\rm e}$) modes spectrum including higher-order terms of $\Big(\frac{\Phi(t)}{M_{P}}\Big)$}}
 \label{appenA}

In the expressions of $H(t),~ H^3(t),~\dot{H}(t)$ and $\ddot{H}(t)$, we truncate the expansions at the order three $\Big(\frac{\Phi(t)}{M_{P}}\Big)^3$ through an amplitude comparison with the other higher-order terms($\mathcal{O}(4),~ \mathcal{O}(5)$ etc.).\\ The expression of the Hubble scale up to $O(4)$ higher-order terms is given as, 
\begin{align}\label{Hubblefornappen}
H(t)\simeq \bar{H}\Bigg(1+\frac{\mathcal{P}\sqrt{6(1-\mathcal{P}^{2n}})}{2(n+1)}\left(\frac{\Phi(t)}{M_{P}}\right)-\frac{3 \mathcal{P}^2}{2(n+1)^2} \left(\frac{\Phi(t)}{M_{P}}\right)^2+\frac{3\sqrt{6}\mathcal{P}^3\sqrt{(1-\mathcal{P}^{2n}})}{4(n+1)^3}\left(\frac{\Phi(t)}{M_{P}}\right)^3\Bigg) 
\end{align}
Subject to this Hubble scale (\ref{Hubblefornappen}), the expressions of $H^3,~ \dot{H},~H\dot{H}$ and $\ddot{H}$ with the higher-order terms can be written as follows 
\begin{align}\label{allHubbleappen}
& H^3(t)\simeq  \bar{H}^3\Bigg(1+\frac{3\mathcal{P}\sqrt{6(1-\mathcal{P}^{2n}})}{2(n+1)}\left(\frac{\Phi(t)}{M_{P}}\right)-\frac{9 \mathcal{P}^2\mathcal{P}^{2n}}{2(n+1)^2} \left(\frac{\Phi(t)}{M_{P}}\right)^2-\frac{3\sqrt{6}\mathcal{P}^3\sqrt{(1-\mathcal{P}^{2n}})}{4(n+1)^3}\left(2+\mathcal{P}^{2n}\right)\left(\frac{\Phi(t)}{M_{P}}\right)^3\Bigg)\nonumber\\
\vspace{20px}
& \dot{H}(t)\simeq 3\bar{H}^2\bigg(\left(\mathcal{P}^{2n}-1\right)-\frac{\sqrt{6}\mathcal{P}\sqrt{1-\mathcal{P}^{2n}}}{(n+1)}\left(\frac{\Phi(t)}{M_{P}}\right)+\frac{9 \mathcal{P}^2}{2(n+1)^2}\left(2\mathcal{P}^{2n}-3\right) \left(\frac{\Phi(t)}{M_{P}}\right)^2\bigg)\nonumber\\
\vspace{20px}
& H(t)\dot{H}(t)\simeq 3\bar{H}^3\bigg(\left(\mathcal{P}^{2n}-1\right)+\frac{\sqrt{6}\mathcal{P}\sqrt{1-\mathcal{P}^{2n}}}{2(n+1)}\left(\mathcal{P}^{2n}-3\right)\left(\frac{\Phi(t)}{M_{P}}\right)+\frac{3 \mathcal{P}^2}{2(n+1)^2}\left(3\mathcal{P}^{2n}-4\right) \left(\frac{\Phi(t)}{M_{P}}\right)^2\nonumber\\
&\quad\quad\quad\quad\quad+\frac{3\sqrt{6} \mathcal{P}^3\sqrt{1-\mathcal{P}^{2n}}}{4(n+1)^3}\left(3\mathcal{P}^{2n}-2\right) \left(\frac{\Phi(t)}{M_{P}}\right)^3\bigg)\nonumber\\
\vspace{20px}
& \ddot{H}(t)\simeq \frac{9 \bar H^3}{(n+1)}\bigg(\left(4-(4+2n){\mathcal P}^{2n}\right)+\sqrt{6}\mathcal{P}\sqrt{1-\mathcal{P}^{2n}}\left(\frac{n+4}{n+1}-\mathcal{P}^{2n}\right)\left(\frac{\Phi(t)}{M_{P}}\right)+\frac{3\mathcal{P}^2}{(n+1)}\left(\frac{3n+4}{n+1}-3\mathcal{P}^{2n}\right)\left(\frac{\Phi(t)}{M_{P}}\right)^2
\nonumber\\
&\quad\quad\quad
+\frac{3\sqrt{6}\mathcal{P}^3\sqrt{1-\mathcal{P}^{2n}}}{(n+1)^2}\left(\frac{\Phi(t)}{M_{P}}\right)^3\bigg)-6\sqrt{6}n\bar{H}^3 \left(\frac{M_{P}}{\Phi(t)}\right)\mathcal{P}^{2n-1}\sqrt{1-\mathcal{P}^{2n}}
\end{align}
\vspace{10px}
Substituting the Equations (\ref{Hubblefornappen}) and (\ref{allHubbleappen}) to (\ref{Eq:mydot}) we obtain
\begin{align}\label{Eq:mydotomegak2appen}
  \frac{\dot{\omega}_k}{\omega_k}\simeq 
  &
  \frac{1}{\big(\frac{k^2}{a^2}+ m^2_{\chi}\big)}
  \bigg[H m^2_{\chi}+
  \left(\frac{7n-11}{n+1} +\frac{9}{n+1}{\cal P}^{2n}
  \right)\bar H^3+\frac{3\sqrt{6}\bar H^3\mathcal{P}\sqrt{1-\mathcal{P}^{2n}}}{2(n+1)^2}(4n-5)\left(\frac{\Phi(t)}{M_{P}}\right)\nonumber\\ 
  &+\underbrace{\frac{9\bar H^3\mathcal{P}^2}{2(n+1)^2}\left(\frac{3n}{n+1}+2\mathcal{P}^{2n}\right)\left(\frac{\Phi(t)}{M_{P}}\right)^2+\frac{3\sqrt{6}\bar H^3\mathcal{P}^3\sqrt{1-\mathcal{P}^{2n}}}{4(n+1)^3}\left(4-25\mathcal{P}^{2n}\right)\left(\frac{\Phi(t)}{M_{P}}\right)^3}_{\text{Additional higher-order terms} }\nonumber\\
  &+ 3\sqrt{6}n \bar H^3\left(\frac{M_{P}}{\Phi(t)}\right)\mathcal{P}^{2n-1}\sqrt{1-\mathcal{P}^{2n}}
 \bigg]  
\end{align}
Similarly, we can define the integral (\ref{Eq:mybeta4}) with the modified expression (\ref{Eq:mydotomegak2appen}) as
\begin{multline}\label{Eq:mybeta4appen}
 \beta_k \simeq 
 \frac{1}{2}\sum_{\nu,l\neq 0}\int_{t_{\text{e}}}^{t}dt^{'}
 \left(\frac{\te}{t'}\right)^3
 \Bigg[ {\cal N}_0e^{i(\nu+l)\omega t^{'}} 
 \left(\frac{t'}{\te}\right)^\frac{1}{n}+ {\cal N}_1 e^{i\nu\omega t^{'}}+{\cal N}_2 +~ \mathcal{N}_3e^{i(\nu+l)\omega t^{'}}\left(\frac{\te}{t}\right)^\frac{1}{n}
 +{\cal N}_4 \left(\frac{t'}{\te}\right)^2+{\cal N}_5e^{i(\nu+l)\omega t^{'}} \left(\frac{t'}{\te}\right)^{\frac{2n-1}{n}}\\+ \underbrace{\mathcal{N}_6e^{i\nu\omega t^{'}}\left(\frac{t^{'}}{\te}\right)^\frac{2(n-1)}{n}+\mathcal{N}_7e^{i(\nu+l)\omega t^{'}}\left(\frac{t^{'}}{\te}\right)^\frac{2n-3}{n}
+\mathcal{N}_8e^{i\nu\omega t^{'}}\left(\frac{\te}{t}\right)^\frac{2}{n}
+\mathcal{N}_9e^{i(\nu+l)\omega t^{'}}\left(\frac{\te}{t}\right)^\frac{3}{n}}_{\text{Additional coefficients}}
\Bigg]
\times \frac{e^{-2i\Omega_k(t^{'})}}{{\left(\frac{k^2}{a^2}+m^2_{\chi}\right)}} \,,
\end{multline}
where newly defined indices are
\bea
  && {\cal N}_6=-{\cal N}_4 \frac{3\mathcal{P}_{\nu}^2}{2(n+1)^2}\left(\frac{\phi_{\rm e}}{M_{P}}\right)^2\,; ~{\cal N}_7={\cal N}_4 \frac{3\sqrt{6}\mathcal{P}_{\nu}^3\left(\sqrt{1-\mathcal{P}^{2n}}\right)_{l}}{4(n+1)^3}\left(\frac{\phi_{\rm e}}{M_{P}}\right)^3\, ; \nonumber\\
  &&{\cal N}_8=\frac{9H_{\rm e}^3}{2(n+1)^2}\left(\frac{3n\mathcal{P}_{\nu}^2}{n+1}+2\mathcal{P}_{\nu}^{2(n+1)}\right)\left(\frac{\phi_{\rm e}}{M_{P}}\right)^2\, ;\nonumber\\
  &&{\cal N}_9=\frac{3\sqrt{6}H_{\rm e}^3}{4(n+1)^3}\left(4\mathcal{P}_{\nu}^3\left(\sqrt{1-\mathcal{P}^{2n}}\right)_l-25\mathcal{P}_{\nu}^{3+2n}\left(\sqrt{1-\mathcal{P}^{2n}}\right)_l\right)\left(\frac{\phi_{\rm e}}{M_{P}}\right)^3
 \label{Eq:niappen}
\eea
Here we introduce a few other separate Fourier series 
having Fourier components $\mathcal{P}^2_{\nu}, \mathcal{P}^3_{\nu}, \mathcal{P}^{2(n+1)}_{\nu}$ and $ \mathcal{P}^{3+2n}_{\nu}$.\\ 

In the range $0\leq w_{\phi}<1/3$, as we have argued in (\ref{beta0.31}), the integrals associated with the coefficients ${\cal N}_1,~ {\cal N}_3$ will not contribute to the stationary points considerably because of the subdominant spectral indices. Likewise, for massless or very low-mass case($m_{\chi}\approx 0$), additional higher-order terms with the coefficients ${\cal N}_8$ and ${\cal N}_9$ have even smaller contributions compared to ${\cal N}_1, {\cal N}_3$. Therefore, in the range $0\leq w_{\phi}<1/3$, the spectral behavior in Eq.(\ref{beta0.3f}) will remain unchanged by adding higher-order terms.\\

In the range $w_{\phi}>1/3$, these additional higher-order terms have a significant contribution to the total amplitude of the spectrum. Evaluating the integral (\ref{Eq:mybeta4appen}) for $w_{\phi}>1/3$ and proceeding along the same way as described in Section \ref{sec4} (see Eqs.(\ref{Iint})-(\ref{smallspectra})) we finally obtain the modified UV spectrum as follows
\begin{equation}\label{smallspectraappen}
\boxed{|\beta_k|^2_{\rm UV}\simeq \frac{1}{16(f(w_{\phi}))^2}\left(\frac{a_{\rm e}}{k} \right)^6 \times \sum\sum\Bigg[ {\cal N}_0 + {\cal N}_1 +{\cal N}_2+{\cal N}_3+{\cal N}_8+{\cal N}_9\Bigg]^2}
\end{equation}
Although the terms associated with ${\cal N}_3, {\cal N}_8, {\cal N}_9$ are faster decaying, while evaluating the integral (\ref{Eq:mybeta4appen}), they have non-negligible contribution in the total amplitude from the perspective of initial time. 

\twocolumngrid
\bibliographystyle{apsrev4-1}
\bibliography{AYANreferences}

\end{document}